\documentclass[range]{ar2e_preprint}
\usepackage{ARAstroBib}

\def\simgt{\lower.5ex\hbox{$\; \buildrel > \over \sim \;$}}
\def\simlt{\lower.5ex\hbox{$\; \buildrel < \over \sim \;$}}

\begin{document}

\input epsf.tex    
\input epsf.def   

\input psfig.sty

\jname{Annu. Rev. Earth Planet. Sci.}
\jyear{2013,}
\jvol{Vol 41:469-95}
\ARinfo{}

\title{The Formation and Dynamics of Super-Earth Planets}

\markboth{Nader Haghighipour}{Super-Earths}

\author{Nader Haghighipour
\affiliation{Institute for Astronomy and NASA Astrobiology Institute, 
University of Hawaii, Honolulu, HI 96822, USA, Email:nader@ifa.hawaii.edu}}


\begin{abstract}
Super-Earths, objects slightly larger than Earth and slightly smaller than
Uranus, have found a special place in exoplanetary science. As a new class of
planetary bodies, these objects have challenged models of planet formation
at both ends of the spectrum and have triggered a great deal of research
on the composition and interior dynamics of rocky planets in connection to
their masses and radii. Being relatively easier to detect than an Earth-sized
planet at 1 AU around a G star, super-Earths have become the focus of
worldwide observational campaigns to search for habitable planets. With a
range of masses that allows these objects to retain moderate atmospheres and
perhaps even plate tectonics, super-Earths may be habitable if they maintain
long-term orbits in the habitable zones of their host stars. Given that in the
past two years a few such potentially habitable super-Earths have in fact been
discovered, it is necessary to develop a deep understanding of the formation
and dynamical evolution of these objects. This article reviews the current
state of research on the formation of super-Earths and discusses different
models of their formation and dynamical evolution.
\end{abstract}

\maketitle
\section{INTRODUCTION}
\label{sec:introduction}
The discovery of planets around other stars has undoubtedly revolutionized our understanding
of the formation and dynamical evolution of planetary systems. The diverse and surprising
characteristics of these objects, both in orbital configuration and physical properties, have
confronted astronomers with many new challenges and have reinvigorated the fields of planet
formation and dynamics.

One surprising characteristic of the currently known extrasolar planets is the range of their
masses. Unlike in the Solar System, where planets belong to two distinct categories of terrestrial
(with masses equal to that of Earth or slightly smaller) and giant [$\sim 14$ Earth masses $({M_\oplus})$ 
and larger], many extrasolar planets have masses in an intermediate range, from slightly larger than
Earth to 10 ${M_\oplus}$. Dubbed super-Earths, these objects present a new class of planetary bodies with
physical and dynamical properties that for the past few years have been the focus of research
among many planetary scientists.

The first super-Earth around a main sequence star was discovered by Rivera et al. (2005)
using the radial velocity technique. [Note that in 1992, Wolszczan \& Frail (1992) discovered
at least two terrestrial-class planets around the pulsar PSR 1257+12.] Thanks to ground-based
observational projects such as the HARPS Search for Southern Extrasolar 
Planets\footnote{http://www.eso.org/sci/facilities/lasilla/instruments/harps/}, 
the California Planet Survey (CPS)\footnote{http://www.exoplanets.org/cps.html}, 
the Lick-Carnegie Exoplanet Survey (LCE), M2K (Clubb et al. 2009), and the MEarth Project 
(Nutzman \& Charbonneau 2008; Irwin et al. 2009a,b)\footnote{http://www.cfa.harvard.edu/MEarth/Welcome.html}, 
and the ongoing success of the CoRoT\footnote{http://smsc.cnes.fr/COROT/index.htm} and 
Kepler\footnote{http://kepler.nasa.gov/}
space telescopes, to date, the number of these objects has exceeded 90. Tables 1 and 2 show the masses and
orbital elements of the currently known super-Earths. As shown, the vast majority of these objects
have orbital periods smaller than 50 days. A survey of the parent stars of these bodies indicates that
more than half of these stars are hosts to multiple planets. This implies that super-Earths may be
more likely to form in short-period orbits and in systems with multiple bodies —- two characteristics
that play important roles in developing models of their formation and dynamical evolution.

Among the currently known super-Earths, a few have gained special attention. CoRoT-7 b, the
seventh planet discovered by the CoRoT space telescope (L\'eger et al. 2009; Queloz et al. 2009;
Hatzes et al. 2010, 2011), and GJ 1214 b, the first super-Earth discovered by transit photometry
around an M star (Charbonneau et al. 2009), are the first super-Earths for which the values of
mass and radius have been measured [CoRoT-7 b: 2.3–8 ${M_\oplus}$, 1.65 Earth radii $({R_\oplus})$; 
GJ 1214 b: 5.69 ${M_\oplus}$, 2.7 ${R_\oplus}$]. This major achievement has enabled theoreticians 
to develop models for the
evolution of super-Earths’ interiors (e.g., Valencia et al. 2006, 2007a,b,c, 2009, 2010; O’Neill \&
Lenardic 2007; Sotin \& Schubert 2009; Tackley \& van Heck 2009) and their possible atmospheric
properties (e.g., Miller-Ricci et al. 2009; Seager \& Deming 2009; Bean et al. 2010; Miller-Ricci
\& Fortney 2010; Rogers \& Seager 2010a,b; Bean et al. 2011; D\'esert et al. 2011; Heng \& Vogt
2011; Berta et al. 2012; Menou 2012; Fraine et al. 2013). The three super-Earth-class bodies GL
581 d (Mayor et al. 2009, Forveille et al. 2011), GL 581 g (Vogt et al. 2010, 2012), and GJ 667C c
(Anglada-Escud\'e et al. 2011) have also made headlines. These planets are the first terrestrial-class
objects that have been discovered in their respective habitable zones.

For the past few years, the formation and characteristics of super-Earths have been the subject
of extensive research.This is primarily because being slightly larger than a typical terrestrial planet,
these objects have the capability of developing moderate atmospheres and may have dynamic interiors
with plate tectonics —- two conditions that would render a super-Earth potentially habitable if
its orbit were in the habitable zone of its host star (see Haghighipour 2011 for a complete review).
Also, unlike Earth-sized planets, super-Earths are relatively easy to detect. Current observations of
super-Earths have indicated that these objects seem to be more common around cool and low-mass
stars (see, e.g., Dressing \& Charbonneau 2013, Swift et al. 2013), where the habitable zone is in
closer orbit. Two prime examples of such systems are GL 581, an M3V star with one or two potentially
habitable super-Earths (Mayor et al. 2009; Vogt et al. 2010, 2012; Forveille et al. 2011), and
the M1.5 star GJ 667C, with a 4.5 ${M_\oplus}$ planet in its habitable zone (Anglada-Escud\'e et al. 2011).

Given the success of observational techniques in detecting potentially habitable super-Earths,
and that during the past two years the number of these objects increased twofold, it would be
natural to expect that many more habitable super-Earths will be detected in the near future. It
is, therefore, imperative to develop a thorough understanding of the formation and dynamical
evolution of these bodies, particularly in connection with their habitability. This article presents
a review of the current state of research on this topic.

Since there are no super-Earths in the Solar System, it is important to know whether the
formation of these objects requires developing new models of planet formation or whether one
can use the models of the formation of planets in the Solar System to explain the formation of super-
Earths. In the latter case, these models will require major revisions. For instance, one characteristic
of super-Earths that presents a challenge to the theories of planet formation is their close-in orbits.
While some models suggest that super-Earths were formed at large distances and migrated to their
present locations, other models present the possibility of their in-place formation. Fortunately, the
physical characteristics of super-Earths, namely their densities,when considered within the context
of different planet formation scenarios, present a potential pathway for differentiating between
these models. In that respect, the study of super-Earths plays an important role in identifying the
most viable planet formation mechanism. The rest of this article presents a review of the current
state of research on this topic.

I begin in Section 2 by briefly reviewing the models of planet formation in the Solar System.
In Section 3, I discuss in detail the application of these models to the formation of super-Earths,
and I conclude in Section 4.

\section{MODELS OF PLANET FORMATION}
\label{sec:FormationModels}

Explaining the formation of planets is one of the most outstanding problems in planetary astronomy.
Despite centuries of efforts to explain the formation of the planets of the Solar System,
this problem is still unresolved, and planet formation is still an open question. The discovery of
extrasolar planets has added even more to these complexities. As explained in Section 1, many of
these objects have physical and orbital properties that are unlike those of the planets in the Solar
System and are not well explained by the current models of Solar System formation and dynamics.

Although the diversity of extrasolar planets has been a continuous challenge to the models of
planet formation, a common practice in explaining the formation of these objects has been to
modify, revise, and/or complement the models of planet formation in the Solar System in such
a way that they would be applicable to other planetary bodies. This suggests that to understand
the formation of extrasolar planets (such as super-Earths), it is necessary to develop a deep understanding
of the models of giant and terrestrial planet formation in the Solar System. This section
is devoted to this task. I begin by explaining the growth of dust particles to larger bodies, then
discuss different phases of planet growth until a full giant or terrestrial planet is formed.

It is widely accepted that planet formation begins in a circumstellar disk of gas and dust known
as a nebula by the growth of dust particles to larger objects. This process, highly dependent on
the mass and dynamical properties of the nebula, proceeds in four stages:

\begin{itemize}
\item coagulation of dust particles through gentle hitting and sticking, which results in the formation
of centimeter- and decimeter-sized objects;

\item growth of centimeter- and decimeter-sized bodies to kilometer-sized planetesimals;

\item collision and accretion of planetesimals to planetary embryos (moon- toMars-sized objects)
in the inner part of the Solar System and to the cores of giant planets in the outer parts; and

\item the accretion of gas and formation of giant planets followed by the collisional growth of
planetary embryos to terrestrial-class bodies.

\end{itemize}

\noindent
The first stage of this process is well understood. Dust grains at this stage undergo different
types of random and systematic motions (Weidenschilling 1977) and frequently collide with one
another. Particles smaller than 100 $\mu$m are mainly subject to Brownian motion and collide with
relative velocities smaller than 1 mm s−1. Larger objects, although slightly faster, are still strongly
coupled to the gas, and their dynamics is governed by the gravitational attraction of the central
star, nongravitational forces such as radiation pressure, and their interaction with the nebula
through gas drag. Gas molecules, however, are subject to pressure gradient (which is necessary
for maintaining the gas at hydrostatic equilibrium), and as a result, their velocities are slightly
smaller than Keplerian. The slight velocity differences between dust particles and gas molecules
cause dust grains to drift inward and approach one another with small relative velocities (Safronov
1969; Weidenschilling 1980; Nakagawa et al. 1981, 1986; Supulver \& Lin 2000; Dullemond \&
Dominik 2005). Turbulence also causes dust grains to collide and is more effective among same-sized
particles. As the collisions of dust particles are gentle, van der Waals forces act between
their surfaces and stick the dust particles to one another. As shown by laboratory experiments and
computational simulations, such gentle collisions result in the fractal growth of dust grains to larger
aggregates (Figure 1) (Smoluchowski 1916; Dominik \& Tielens 1997; Blum et al. 1998; Wurm
\& Blum 1998; Blum \& Wurm 2000; Krause \& Blum 2004; Blum 2006, 2010; Wada et al. 2007).

While the process of the growth of micrometer-sized dust grains to millimeter- and centimeter-sized
objects is well understood, the growth of the latter bodies to larger sizes (i.e., kilometer size)
is still a big mystery. Simulations have shown that as dust particles grow, their coupling to the
gas weakens (i.e., their velocities relative to the gas molecules increase), and they show more of
their independent dynamics (Weidenschilling 1977). At this stage, differential vertical settling
(Safronov 1969), radial drift (Whipple 1972), and turbulence (V\"olk et al. 1980, Mizuno et al.
1988, Ormel \& Cuzzi 2007) play important roles in driving particles’ relative velocities. The
latter causes objects to approach each other rapidly and increases their impact velocities. Results
of laboratory experiments and computational simulations have shown that as objects grow to
centimeters in size, their sticking efficiency drops dramatically (Blum \& M\"unch 1993), and their
relative velocities become so large that their collisions may result in bouncing (bouncing barrier)
and/or erosion and fragmentation (fragmentation barrier) (Blum \& Wurm 2008, G\"uttler et al.
2009, Zsom et al. 2010, Beitz et al. 2011).

The above-mentioned bouncing and fragmentation barriers are not the only obstacles in the
formation of planetesimals. The sub-Keplerian rotational velocities of gas molecules result in the
transfer of angular momentum from solid bodies to the gas and the subsequent drift of these objects
toward the central star. The rate of this radial drift is approximately proportional to the size of an
object, implying that as an object grows, it approaches the central star in a shorter time. Numerical
simulations have indicated that meter-sized bodies have the fastest radial drifts. Combined with
turbulence and differential settling, this radial drift increases the relative velocities of solid objects
and causes many of them to collide with one another at large speeds. Given that large objects
are more prone to collisional destruction (the sticking properties of solid bodies weaken as they
grow), it is expected that many of these impacts result in the breakage of the colliding bodies.
This process, known as the meter-size barrier, implies that even if the centimeter-size bouncing
barrier is overcome, the impact velocities of solid objects become so large that their collisions
result in their breaking into small fragments, which subsequently halts their growth to larger
sizes. These fragments, even if reaccumulated, will go through the same above-mentioned process
and ultimately drift into the central star, leaving the nebula devoid of the solid material necessary
for the formation of planetesimals.

Interestingly, despite all these difficulties, planets do exist and so do many kilometer-sized
bodies, such as the asteroids and Kuiper belt objects. This implies that during the early stages
of planet formation, Nature succeeded in finding a way to overcome the centimeter-sized and
meter-sized barriers. It may be that kilometer-sized planetesimals did not form as a result of the
mere collisional growth of dust grains; other mechanisms may have also contributed.

A planet-forming nebula is a dynamic environment whose properties and structure vary with
time. These variations, in particular in a gaseous disk, may manifest themselves as different structures
in the nebula. For instance, regions may appear where the pressure of the gas is locally
enhanced. The appearance of such structures will immediately affect the motions of particles in
their surroundings. As opposed to a nebula with a monotonic radial pressure profile where gas
drag and pressure gradient cause inward migration of solids, in the vicinity of pressure-enhanced
regions, the velocity differences between solid objects and gas molecules cause solid particles
to undergo inward and outward migrations and to accumulate around the locations of pressure
maxima (Haghighipour \& Boss 2003a,b; Haghighipour 2005).

In a gaseous disk, the turbulent eddies created by magnetorotational instability are examples
of such high-pressure regions. As Johansen et al. (2006, 2007, 2008) have shown, the formation
of these turbulent eddies causes small centimeter- and decimeter-sized objects to accumulate
in their vicinities and increases the local density of solid material. As the accumulation of solid
objects continues, their local spatial density increases until their region becomes gravitationally
unstable and the accumulated bodies fragment into several 100–1,000-km-sized planetesimals.
This mechanism, known as streaming instability, has been presented as a scenario for planetesimal
formation. [See Chiang \& Youdin (2010) for a review and Cuzzi et al. (2008) and Weidenschilling
(2010) for alternative viewpoints.]

It is important to note that as shown by Shariff \& Cuzzi (2011), the local enhancement of solid
to gas surface density necessary for the onset of instability is achievable only when the turbulence
is extremely weak. These authors indicate that when the effect of turbulent mass diffusivity is
taken into account, streaming instability becomes inefficient, and the growth rate of planetesimals
reduces significantly.

Other mechanisms of the formation of planetesimals include trapping dust particles in vortices
(Barge \& Sommeria 1995, Klahr \& Henning 1997, Lyra et al. 2009a), trapping particles in pressure
enhanced regions created by the evaporation front of water in the protoplanetary disk (Kretke \&
Lin 2007; Brauer et al. 2008a,b; Lyra et al. 2009b), turbulent concentration of solids (Chambers
2010), turbulent clustering of protoplanetary bodies (Pan et al. 2011), concentration of solid objects
at the snowline (the region beyond which water is in the permanent state of ice) as a result of the
sublimation of drifting ice aggregates (Aumatell \& Wurm 2011), trapping of solid objects in dead
zones (Gressel et al. 2012) and at the boundary between steady super/sub-Keplerian flow created
by inhomogeneous growth of magnetorotational instabilities (Kato et al. 2012), rapid coagulation
of porous dust aggregates outside the snowline (Okuzumi et al. 2012), and planetesimal formation
in self-gravitating disks (Gibbons et al. 2012, Shi \& Chiang 2013).

The four stages of planet formation outlined above share one interesting feature: The underlying
physics of each stage is almost distinct from that of the other phases. This makes it possible to
study each phase separately. Once the dust grains have grown and kilometer-sized planetesimals
are formed, although the circumstellar disk still contains gas and dust, its dynamics is now mainly
driven by the interaction of planetesimals with one another. These interactions are primarily
gravitational, although gas drag also plays a role. At this stage, because the planetesimals are the
main components populating the disk, collisions among these objects are frequent, which results
in low eccentricities and low inclinations for these bodies. Because the relative velocity between
two bodies is an increasing function of their orbital eccentricities, lowering the eccentricity of
planetesimals due to their mutual collisions and dynamical friction, combined with their almost
coplanar orbits, reduces their relative velocities. The latter facilitates the merging of these objects
and enhances the rate of their accretion to larger bodies.

As a planetesimal grows, the influence zone of its gravitational field expands and as a result,
it attracts more material from its surroundings. In other words, more material will be available
for the planetesimal to accrete, and the rate of its growth increases. Known as runaway growth,
this process results in the growth of kilometer-sized planetesimals to larger bodies in a short time
(Safronov 1969; Greenberg et al. 1978; Wetherill \& Stewart 1989, 1993; Ida \& Makino 1993;
Kokubo \& Ida 1996, 2000; Weidenschilling et al. 1997).

Runaway growth is a local process. Since the collision of two objects is more likely to result in
their coalescence when their relative velocity is small, the effectiveness of this process in producing
larger bodies, and the type and size of the resulting objects, varies at different distances from the
central star. At large distances (e.g., $> 5$ AU from the Sun), where the rotational velocities are small,
planetesimals approach each other with small relative velocities, and their impacts are likely to
result in accretion. Also, because the temperature in the circumstellar disk is low at such distances,
the bulk material of such planetesimals is primarily ice, which increases the efficiency of their
sticking at the time of their collision. As a result, planetesimals at large distances grow to objects
of a few Earth masses in a short time. As this process occurs while the nebular gas is still present,
a growing object gradually attracts gas from its surroundings, forming a large body with a thick
gaseous envelope and a mass equal to a few hundred Earth masses. At this state, a gas-giant planet
is formed. This scenario, known as the core-accretion model, has been proposed as a mechanism
for the formation of gas-giant planets in the Solar System (Pollack et al. 1996, Hubickyj et al.
2005, Lissauer et al. 2009, Movshovitz et al. 2010).

As the giant planets form at large orbits, the runaway accretion takes a slightly different path in
the inner parts of the disk. Similar to the formation of the cores of gas-giant planets, the collisions
of planetesimals at this stage may result in their growth to larger bodies. However, because the
orbital motions of planetesimals are faster, they may approach each other with larger relative
velocities. Also, many of these objects may lose their surface ices and other volatiles at closer
distances, and as a result, when they collide with one another, the efficiency of their accretion will
not be as high as for those at larger orbits. Simulations of the collision and growth of planetesimals
in the inner part of the Solar System have shown that instead of forming objects as big as the cores
of giant planets, accretion of these bodies results in the formation of several hundred moon- to
Mars-sized objects known as planetary embryos. Computational simulations (Bromley \& Kenyon
2006) and analytical analysis (Goldreich et al. 2004) have shown that when the masses of these
embryos reach lunar mass, the dynamical friction of the swarm of planetesimals can no longer
dampen their orbits, and their runaway growth ends. At this stage, the gravitational perturbation of
the resulting planetary embryos, combined with the perturbation of giant planets, strongly affects
the dynamics of smaller planetesimals and causes many of them to collide at high velocities and
shatter one another, and/or their orbits become highly eccentric, and they subsequently scatter to
large distances where they may leave the gravitational field of the system. This growth and clearing
process continues until terrestrial planets are formed and the smaller remaining bodies (asteroids)
are in stable orbits (Figure 2) (Wetherill 1990a,b, 1994, 1996; Kokubo \& Ida 1995, 1998, 2007;
Chambers \& Wetherill 1998, 2001; Agnor et al. 1999; Morbidelli et al. 2000, 2012; Chambers
2001; Chambers \& Cassen 2002; Levison \& Agnor 2003; Raymond et al. 2004, 2005a,b, 2006b,
2007, 2009; Kokubo et al. 2006; O’Brien et al. 2006; Hansen 2009; Schlichting et al. 2012; Torres
et al. 2013; Haghighipour et al. submitted; Izidoro et al. submitted). 
Since the accretion and reaccretion of bodies in smaller
orbits are not as efficient as in the outer regions, unlike the growth of gas-giant planets, the
formation of terrestrial bodies will take several hundred million years. Figure 2 shows the time
evolution of a sample simulation of terrestrial planet formation (Haghighipour et al. submitted; 
Izidoro et al. submitted).
The planet formation models as explained above, although capable of explaining many features
of the Solar System, face several complicated challenges. The core-accretion model, for instance,
requires the nebular gas to be available for $\sim 10$ Ma while the core of Jupiter grows and accretes gas
from its surroundings (Pollack et al. 1996). However, the observational estimates of the lifetimes
of disks around young stars suggest a lifetime of 0.1–10 Ma, with 3 Ma being the age at which
half the stars show evidence of disks (Strom et al. 1993, Haisch et al. 2001, Chen \& Kamp 2004,
Maercker et al. 2006). These simulations also suggest a solid core for Jupiter with a mass of
$\sim 10 {M_\oplus}$. Computational modeling of the interiors of Jupiter and Saturn, however, has indicated
different possible values for the cores of these objects, ranging from 0 to as large as $14 {M_\oplus}$ 
(Guillot 2005, Militzer et al. 2008). It is unclear what the actual masses of the cores of our gas-giant planets
are, and if smaller than $10 {M_\oplus}$, how they accumulated their thick envelopes in a short time. I refer
the reader to a review by Guillot (2005) for more details.

To overcome these difficulties, the core-accretion model has undergone several improvements.
Hubickyj et al. (2005) and Lissauer et al. (2009) have shown that increasing the surface density of
the nebula to higher than that suggested by Pollack et al. (1996) significantly reduces the time of
the giant planet formation. An improved treatment of grain physics as given by Podolak (2003),
Movshovitz \& Podolak (2008), and Movshovitz et al. (2010) has also indicated that the value of
the grain opacity in the envelope of the growing Jupiter in the original core-accretion model
(Pollack et al. 1996) is too high, and a lower value has to be adopted. This lower opacity has
led to a revised version of the core-accretion model in which the time of giant planet formation
is considerably smaller (Hubickyj et al. 2005, Movshovitz et al. 2010). Most recently, Bromley
\& Kenyon (2011) have developed a new hybrid N-body-coagulation code that has enabled the
authors to form Saturn- and Jupiter-sized planets in $\sim 1$ Ma.

An alternative model for the formation of gas-giant planets addresses this issue by proposing
rapid formation of giant planets in a gravitationally unstable nebula (Boss 2000a,b, 2003; Mayer
et al. 2002, 2004, 2007; Durisen et al. 2007; Boley 2009; Boley et al. 2010; Cai et al. 2010). Known
as the disk-instability scenario, this model suggests that local gravitational instabilities in the solar
nebula may result in the fragmentation of the disk to massive clumps that subsequently contract
and form gas-giant planets in a short time. Boss's (2000a,b) and Mayer et al.'s (2002, 2003, 2004)
results show that an unstable disk can break up into giant gaseous protoplanets in as short a time
as $\sim 1,000$ years. Although this mechanism presents a fast track to the formation of a gas-giant
planet, it suffers from the lack of an efficient cooling process necessary to take energy away from
a planet-forming clump in a sufficiently short time before it disperses.

\section{FORMATION OF SUPER-EARTHS}
\label{Super-EarthFormation}

The extent to which current planet formation scenarios can be used to explain the formation of
super-Earths varies with the mass and orbital architecture of these objects. Since the dynamics
and characteristics of planet-forming nebulae are different for stars with different spectral types,
the parent stars of super-Earths also play an important role. The range of masses for the currently
known super-Earths, when considered within the context of giant and terrestrial planet formation
scenarios, points to two general pathways for the formation of these objects. The low-mass super-Earths 
could have formed in place following a similar process as the formation of terrestrial
planets in the Solar System (see, e.g., Chiang \& Laughlin 2012). The larger super-Earths, with
masses close to their upper limit, may be the result of an unsuccessful and incomplete giant planet
formation (see, e.g., Rogers et al. 2011). In this scenario, the super-Earths’ larger than terrestrial
masses, combined with the fact that many of these objects are in short-period orbits, point to
a formation scenario in which super-Earths are formed at large distances (where more material
is available for their growth) and either migrate to their current locations as they interact with
the protoplanetary disk (Kennedy \& Kenyon 2008b) or are scattered to their current orbits as
a result of interactions with other cores and/or planets (Terquem \& Papaloizou 2007). In other
words, the formation of these objects may have occurred while their orbital elements were evolving
(Terquem \& Papaloizou 2007; Kennedy \& Kenyon 2008a,b). This mechanism naturally favors
the core-accretion model of gas-giant planet formation, although attempts have also been made
to explain the formation of super-Earths via the disk-instability scenario (see Section 3.3).
As mentioned above, super-Earths owe their popularity to their masses and sizes, which under
favorable conditions may render them habitable. While planet formation models allow for the
formation of super-Earths around all types of stars (either as a failed core of a giant planet or as a
slightly larger terrestrial-class object), because of the current sensitivity of detection techniques, a
great deal of interest exists in super-Earths in the habitable zones of cool and low-mass stars (e.g.,
M dwarfs). For this reason, I devote the rest of this article to presenting a review of the models of
super-Earth formation around M stars.

\subsection{Formation of Super-Earths Around Low-Mass Stars} 

The discovery of planets of different sizes, from Jovian-type [e.g., GJ 876 b, c, and e (Rivera
et al. 2010); HIP 57050 b (Haghighipour et al. 2010); GL 581 b (Bonfils et al. 2005); KOI-254 b
(Johnson et al. 2012); Kepler-32 d (Swift et al. 2013)] to small super-Earths [e.g., GL 581 c, d, e,
and g (Udry et al. 2007, Mayor et al. 2009, Vogt et al. 2010); GJ 667C c (Anglada-Escud\'e et al.
2011); Kepler-32 b and c (Swift et al. 2013)] around M dwarfs indicates that both giant and
terrestrial planet formation can proceed efficiently around low-mass stars. This implies that the
circumstellar disks around these stars can accommodate the formation of super-Earths both as
a failed core of a giant planet through the gas-giant planet formation process, and also as small
terrestrial-class objects through direct collisional growth of protoplanetary bodies and planetary
embryos. These mechanisms have to also account for the short periods of super-Earths, whether
through planet migration, planet-planet scattering, or a combination of both.
I begin this section by considering the core-accretion model as the mechanism for the formation
of super-Earths. As mentioned above, the discovery of super-Earths can be taken as strong evidence
in support of this model. However, as is explained at the end of the next section, this mechanism
alone cannot explain the formation and orbital architecture of all the currently known super-
Earths. Other effects such as the evolution of the central star and planet migration have to be
taken into consideration as well. I discuss these effects in the next section and conclude this article
by reviewing the formation of super-Earths through the disk-instability model.

\subsection{The Core-Accretion Model}

As mentioned in Section 2, the efficiency of the core-accretion model and the rate of the growth
of the cores of giant planets increase with the disk surface density. Around low-mass stars, where
the surface density of the disk is smaller than around the Sun, the solid material (i.e., the planetesimals)
is more spatially scattered, and as a result, the collisions among planetesimals and planetary
embryos are less frequent. This smaller rate of collision prolongs the growth of planetesimals to
larger sizes, and causes the time of the core growth around low-mass stars to be several times
longer than the time of the formation of Jupiter around the Sun. As shown by Laughlin et al.
(2004), in disks around stars with masses smaller than 0.5 solar masses $({M_\odot})$, the core-accretion
mechanism can produce planets ranging from terrestrial-class to Neptune sizes. However, the
time for the formation of these objects is much longer than the time for the formation of Jupiter
in the Solar System through the core-accretion model. During this time, around M stars, for
instance, the gaseous component of the circumstellar disk disperses, leaving the slowly growing
core with much less gas to accrete.

The short lifetime of the gas in circumstellar disks around M stars can be attributed to two
important factors:

\begin{itemize}
\item the high internal radiation of young M stars (at this stage, these stars are almost as bright as
Sun-like stars), and

\item external perturbations from other close-by stars.
\end{itemize}

\noindent
The latter is primarily due to the fact that most stars are formed in clusters (Lada \& Lada 2003), and
as such, their circumstellar disks are strongly affected by the gravitational perturbations and the
radiations of other stars (Adams et al. 2004). For M stars, this causes the circumstellar disk to receive
a high amount of radiation from both the central star and external sources. This high amount of
radiation combined with the low masses of M stars, which points to their small gravitational fields,
increases the effectiveness of the photoevaporation of the gaseous component of the circumstellar
disk by up to two orders of magnitude. As a result, the majority of the gas leaves the disk at the
early stages of giant planet formation, leaving a still-forming core with not much gas to accrete.

\subsubsection{Effect of stellar evolution}

Although the growth of giant planets’ cores through collision
and accretion of planetesimals is similar in disks around solar-type and low-mass stars, the fact that
around smaller stars this process takes longer introduces a fundamental difference in the formation
of giant planets in these two environments. As opposed to young Sun-like stars whose luminosities
stay almost constant during the formation of giant and terrestrial planets (e.g., 10-100 Ma), the
luminosity of a premain sequence, low-mass star (e.g., $0.5 {M_\odot}$) fades by a factor of 10 to 100
during this process (Hayashi 1981). This causes the internal temperature of the circumstellar disk
to decrease, which subsequently causes the disk’s snowline to move toward the central star and to
close distances. The forward migration of the snowline results in an increase in the population of
icy materials (kilometer-sized and larger planetesimals) in the outer regions of the disk, which in
turn increases the efficiency of the collisional growth of these objects to protoplanetary bodies (as
mentioned in Section 2, sticking is more efficient among icy bodies). As shown by Kennedy et al.
(2006), around a 0.25-$M_\odot$ star, the moving snowline causes rapid formation of planetary embryos
within a few million years (also see Kennedy et al. 2007). Subsequent collisions and interactions
among these objects result in the formation of super-Earths in approximately 50-500 Ma.

\subsubsection{Effect of planet migration}

As mentioned above, one of the major developments in the
field of planetary dynamics that was a direct consequence of the detection of extrasolar planets is
the concept of planet migration. Although previously post-formation migration had been proposed
as a mechanism to explain the orbital architecture of small bodies in the Solar System (e.g., moons
of giant planets and Kuiper belt objects), the migration of planets during their formation had not
been incorporated into the models of planetary formation. In other words, the planet formation
scenarios mentioned above were developed assuming that planets form in place.
The discovery of extrasolar planets, almost from the beginning, challenged this assumption.
The detection of the first hot Jupiter in a 4-day orbit around the star 51 Pegasi (Mayor \& Queloz
1995) revealed that planet migration is an inseparable part of the evolution of a planetary system
and prompted astronomers to revisit this concept and to incorporate it into their models of
planet formation. Today, planet migration is well developed and widely accepted as part of a
comprehensive planet formation scenario.

Planetary and satellite migration has long been recognized as a major contributor to the formation
and orbital architecture of planets, their moons, and other minor bodies in the Solar
System. As shown by Greenberg et al. (1972) and Greenberg (1973), mean-motion resonances
(i.e., commensurable orbital periods\footnote{Orbital commensurability is necessary for two planets 
to be in a mean-motion resonance; however, it is not sufficient. Other constraints have to exist between 
the angular elements of their orbits as well. For more details, the reader is referred to Roy
(1982), Danby (1992), and Murray \& Dermott (1999).}) among the natural satellites of giant planets 
(e.g., Titan and Hyperion, satellites of Saturn) may have been the result of the radial migration of these 
objects due to their tidal interactions with their parent planets (Goldreich 1965). The dynamical architecture
of Galilean satellites, with their three-body, Laplace resonance, has also been attributed
to the migration of these objects. It is accepted that these satellites migrated inward during their
formation as a result of interacting with the circumplanetary disk of satellitesimals around Jupiter
(Canup \& Ward 2002), and subsequently by tidal forces after their formation (Peale \& Lee 2002).
The lack of irregular satellites between Callisto, the outermost Galilean satellite, and Themisto,
the innermost irregular satellite of Jupiter, also can be explained by a dynamical clearing process
that occurred during the formation and migration of Galilean satellites (Haghighipour \& Jewitt
2008). Among the planets of our Solar System, the post-formation, planetesimal-driven migration
of giant planets has been proposed as a mechanism to explain the current state of the asteroid belt
(Tsiganis et al. 2005; Minton \& Malhotra 2009, 2011; see also Gomes 1997), late heavy bombardment
(Gomes et al. 2005), the origin of Jupiter Trojan asteroids (Morbidelli et al. 2005), the
effects of secular resonances on terrestrial planet formation (Agnor \& Lin 2012), and the small
mass and size of Mars (Walsh et al. 2011). I refer the reader to Morbidelli et al. (2012) for a review
on these topics.

The idea of the migration of planetary bodies was first proposed by Fernandez \& Ip (1984).
These authors suggested that after the dispersal of the nebular gas, fully formed giant planets
may drift from their original orbits due to the exchange of angular momentum with the disk
of planetesimals. As a result of this post-formation migration, small bodies either are scattered
out of the Solar System or may reach other regions where they may reside in long-term stable
orbits. As shown by Malhotra (1993, 1995), this mechanism can explain the peculiar orbit of Pluto
(highly eccentric, inclined, and long-term chaotic), and as shown by Malhotra (1996) and Hahn
\& Malhotra (2005), it can also explain the dynamical structure of Kuiper belt objects.

The past two decades have witnessed major developments in the theories of planet migration.
Simulations of the formation of planetary bodies and their interactions with circumstellar disks
have shown that planet migration does not have to occur necessarily after the planets are fully
formed. In fact, planets can migrate while they are forming as a result of exchanging angular
momentum with their surrounding environment. This naturally suggests that the physical and
dynamical characteristics of a planet and its circumstellar disk will play an important role in this
process. For instance, the planet may undergo type I migration, in which case it does not accrete
nebular material as it migrates (Figure 3a). Conversely, the planet may be large and accrete
nebular material, in which case it may create a gap in the disk as it migrates (Figure 3b). This
type of migration is known as type II migration. Planet migration may occur in other forms as
well.\footnote{I do not discuss these mechanisms here, as they may not be entirely relevant to the 
formation and dynamical evolution of super-Earths. Instead, I refer the reader to numerous articles 
that have been published on these subjects. Unfortunately, the richness of the literature does not 
allow me to cite all these articles here, but among them, one can refer to Nelson et al. (2001), 
Mass\'et \& Snellgrove (2001), Papaloizou \& Terquem (2006), Chambers (2009), Armitage (2010), 
and a recent review by Baruteau \& Mass\'et (2013).}

The contribution of planet migration to the formation of close-in super-Earths may appear
in different forms. The most common scenario involves the inward migration of a fully formed
giant planet in a disk of planetesimals and planetary embryos. The giant planet in this scenario
affects the dynamics of protoplanetary bodies interior to its orbit by either increasing their orbital
eccentricities and scattering them to larger distances or causing them to migrate to closer orbits.
The migrating protoplanets may be shepherded by the giant planet into small close-in regions,
where they are captured in mean-motion resonances. As Zhou et al. (2005), Fogg \& Nelson
(2005, 2006, 2007a,b, 2009), and Raymond et al. (2008) have shown, around Sun-like stars, the
shepherded protoplanets may also collide and grow to terrestrial-class and super-Earth objects
(see, e.g., Figure 6b). Studies of the back-scattered objects in the simulations of disks around
massive stars have shown that these bodies may also collide and grow to planetary sizes (Mandell
\& Sigurdsson 2003, Raymond et al. 2006a, Mandell et al. 2007).

While around Sun-like stars, despite the out-scattering of protoplanetary bodies during the
migration of a giant planet, the formation of super-Earths through the collision and growth of
planetesimals and planetary embryos proceeds efficiently, around low-mass stars this scenario is
not always the case. Simulations of the dynamics of protoplanetary bodies at distances smaller
than 0.2 AU around a 0.3 ${M_\odot}$ star have shown that during the inward migration of one or several
giant planets (the latter involves migrating planets in mean-motion resonances), the majority of
the protoplanets leave the system and do not contribute to the formation of close-in Earth-sized
bodies and/or super-Earths (Figure 4) (Haghighipour \& Rastegar 2011). These results suggest
that the currently known small planets around M stars might have formed at larger distances and
were either scattered to their current close-in orbits (e.g., GJ 876 d; see Figure 5) or migrated
into their orbits while captured in a mean-motion resonance with a migrating planet.

The above-mentioned scenario for the formation of close-in super-Earths is based on the
fact that giant planets are formed long before the protoplanetary bodies grow to larger sizes.
The underlying assumption in this scenario is that the giant planet does not migrate during its
formation, and the migration of the planetary embryos (the moon- to Mars-sized objects) is also
ignored. However, not only do the cores of still-forming giant planets migrate (Alibert et al. 2004),
so too do the planetary embryos. While migrating, the embryos may undergo orbital crossing and
collisional merging, which may result in their growth to a few super-Earths, especially in mean motion
resonances. Simulating the interactions of 25 protoplanetary objects with masses ranging
from 0.1 to 1 $M_\oplus$, Terquem \& Papaloizou (2007) have shown that a few close-in super-Earths
may form in this way with masses up to $12 {M_\oplus}$. The results of these simulations suggest that
in systems in which merging of migrating cores results in the formation of super-Earths and
Neptune-like planets, such planets will always be accompanied by giant bodies and most likely
will be in mean-motion resonances. Similar results have also been reported by Haghighipour \&
Rastegar (2011).

Interestingly, several planetary systems have been discovered in which central stars host only
small Neptune-sized objects and super-Earths (e.g., HD 69830, GL 581). The planets in these
systems do not have a Jupiter-like companion that could have migrated to facilitate their formation.
Such systems seem to imply that a different mechanism may be responsible for the formation
of their super-Earth bodies. Kennedy \& Kenyon (2008a) and Kenyon \& Bromley (2009) have
suggested that the migration of protoplanetary embryos may be the key in facilitating the close-in
accretion of these objects. These authors considered a circumstellar disk with a density enhancement
at the region of its snowline and simulated the dynamics and growth of its planetary embryos.
They showed that while interacting with one another (colliding and accreting), many of these objects
may migrate toward the central star. Around a solar-type star, the time of such migrations
for an Earth-sized planet at 1 AU is $\sim {10^5} - {10^6}$ years - much smaller than the time for the chaotic
growth of a typical moon- or Mars-sized embryo ($10^8$ years) (Goldreich et al. 2004). This implies
that most of the migration occurs prior to the onset of the final growth. Depending on their relative
velocities, the interactions among the migrating embryos may result in their growth, scattering,
and/or shepherding, as in the case of a migrating giant planet. Simulations by Kennedy \& Kenyon
(2008b) and Kenyon \& Bromley (2009) have shown that super-Earth objects with masses up to
8 ${M_\oplus}$ may form in this way around stars ranging from 0.25 to 2 $M_\odot$ (Figure 6).

\subsection{The Disk-Instability Model}
\vskip -5pt
The formation of super-Earths through the mechanisms explained above, particularly when those
mechanisms are used to explain the formation of these objects at the higher end of their mass
range, naturally favors the core-accretion model of giant planet formation. However, the fact that
Jovian-type planets have been discovered around low-mass stars (e.g., GJ 876, with three planets
ranging from 1 Uranus mass to 2.2 Jupiter masses in $\sim$120-, 60-, and 30-day orbits; HIP57050, with
a Saturn-mass planet in a $\sim$40-day orbit) suggests that the disk-instability model may also be able
to form close-in super-Earths, especially those that are considered as failed cores of giant planets.
As explained above, given the low masses of the circumstellar disks around M stars, the existence
of giant planets around these stars suggests that they might have formed at large distances and
migrated to their current orbits. This is because in a planet-forming nebula, more nebular material
is available at outer regions that can then facilitate the formation of a giant planet through the
core-accretion model. The availability of more mass at outer distances in a disk may also trigger
the formation of giant planets around M stars through the disk-instability scenario. Recall that
in this scenario, clumps, formed in an unstable gaseous disk, collapse and form gas-giant planets
(e.g., Boss 2000b, Mayer et al. 2002). After the giant planets are formed, a secondary process is
needed to remove their gaseous envelopes. As Boss (2006) has shown, such collapsing clumps
can form around a 0.5 $M_\odot$ star at a distance of $\sim$8 AU (Figure 7). This author suggests that, as
most stars are formed in clusters and in high-mass, star-forming regions, intense far/extreme UV
radiations from nearby O stars may rapidly (within 1 Ma) photoevaporate the gaseous envelopes
around giant planets, leaving them with large super-Earth cores. Similar mechanisms have been
suggested for the formation of Uranus and Neptune in the Solar System (Boss et al. 2002). A
subsequent migration, similar to that suggested by Michael et al. (2011), may then move these
cores to close-in orbits.

\section{CONCLUDING REMARKS}
\vskip -5pt
As evident from this review, it is generally accepted that super-Earths are formed through a
combination of a core accumulation process and planetary migration. Modeling the formation
of these objects requires the simulation of the collisional growth of planetary embryos and their
subsequent interactions with the protoplanetary disk. A realistic model requires global treatment
of the disk and inclusion of large numbers of planetesimals and planetary embryos. In practice,
such simulations are computationally expensive. To avoid such complications, most of the current
models of super-Earth formation include only small numbers of objects (e.g., cores, progenitors,
protoplanets, planetesimals). As shown by McNeil \& Nelson (2010), in systems with large numbers
of bodies (e.g., several thousand planetesimals and larger objects), the combination of traditional
core accretion and type I planet migration may not produce objects larger than 3-4 $M_\oplus$ in close-in
(e.g., $\leq 0.5$ AU) orbits. Although the systems studied carry some simplifying assumption, McNeil
\& Nelson’s results point to an interesting conclusion: While the combination of core accretion and
planet migration seems to be a viable mechanism for the formation of close-in super-Earths, the
formation of these objects is still an open question, and a comprehensive theory for their formation
requires more sophisticated computational modeling, with possibly entirely new physics, as yet to
be discovered.

\section{ACKNOWLEDGMENT}
I am grateful to J\"orgen Blum, Alan Boss, Andre Izidoro, Scott Kenyon, Fr\'ed\'eric Mass\'et, 
and Ji-Lin Zhou for kindly providing figures.


\section{REFERENCES}

\vskip 2pt
\noindent
Adams FC, Hollenbach D, Laughlin G, GortiU. 2004. Photoevaporation of circumstellar disks due to external
far-ultraviolet radiation in stellar aggregates. Astrophys. J. 611:360-79

\hfill
\vskip 0.5pt
\noindent
Agnor CB, Canup RM, Levison HF. 1999. On the character and consequences of large impacts in the late
stage of terrestrial planet formation. Icarus 142:219-37

\hfill
\vskip 0.5pt
\noindent
Agnor CB, Lin DNC. 2012. On the migration of Jupiter and Saturn: constraints from linear models of secular
resonant coupling with the terrestrial planets. Astrophys. J. 745:143

\hfill
\vskip 0.5pt
\noindent
Alibert Y, Mordasini C, Benz W. 2004. Migration and giant planet formation. Astron. Astrophys. 417:L25-28

\hfill
\vskip 0.5pt
\noindent
Anglada-Escud\'e G, Arriagad P, Vogt SS, Rivera E, Butler RP. 2011. A planetary system around the nearby
M dwarf GJ 667C with at least one super-Earth in its habitable zone. Astrophys. J. 751:L16

\hfill
\vskip 0.5pt
\noindent
Armitage PJ. 2010. The early evolution of planetary systems. In Astrophysics of Planet Formation, pp. 218-62.
Cambridge, UK: Cambridge Univ. Press

\hfill
\vskip 0.5pt
\noindent
Aumatell G, Wurm G. 2011. Breaking the ice: planetesimal formation at the snowline. MNRAS 418:L1-5

\hfill
\vskip 0.5pt
\noindent
Barge P, Sommeria J. 1995. Did planet formation begin inside persistent gaseous vortices? Astron. Astrophys.
295:L1-4

\hfill
\vskip 0.5pt
\noindent
Baruteau C, Mass´et F. 2013. Recent developments in planet migration theory. In Tides in Astronomy and
Astrophysics (Lecture Notes in Physics), ed. J Souchay, S Mathis, T Tokieda, p. 201. Berlin: Springer-Verlag

\hfill
\vskip 0.5pt
\noindent
Bean JL, D\'esert J-M, Kabath P, Stalder B, Seager S, et al. 2011. The optical and near-infrared transmission
spectrum of the super-Earth GJ 1214b: further evidence for a metal-rich atmosphere. Astrophys. J. 743:92

\hfill
\vskip 0.5pt
\noindent
Bean JL, Miller-Ricci Kempton E, Homeier D. 2010. A ground-based transmission spectrum of the super-Earth 
exoplanet GJ 1214b. Nature 468:669-72

\hfill
\vskip 0.5pt
\noindent
Beitz E, G\"uttler C, Blum J, Meisner T, Teiser J,Wurm G. 2011. Low-velocity collisions of centimeter-sized
dust aggregates. Astrophys. J. 736:34

\hfill
\vskip 0.5pt
\noindent
Berta ZK, Charbonneau D, D\'esert J-M, Miller-Ricci Kempton E, McCullough PR, et al. 2012. The flat
transmission spectrum of the super-Earth GJ1214b from Wide Field Camera 3 on the Hubble Space
Telescope. Astrophys. J. 747:35

\hfill
\vskip 0.5pt
\noindent
Blum J. 2006. Dust agglomeration. Adv. Phys. 55:881-947

\hfill
\vskip 0.5pt
\noindent
Blum J. 2010. Dust growth in protoplanetary disks—a comprehensive experimental/theoretical approach. Res.
Astron. Astrophys. 10:1199-214

\hfill
\vskip 0.5pt
\noindent
Blum J, M\"unch M. 1993. Experimental investigations on aggregate-aggregate collisions in the early solar
nebula. Icarus 106:151-67

\hfill
\vskip 0.5pt
\noindent
Blum J, Wurm G. 2000. Experiments on sticking, restructuring, and fragmentation of preplanetary dust
aggregates. Icarus 143:138-46

\hfill
\vskip 0.5pt
\noindent
Blum J, Wurm G. 2008. The growth mechanisms of macroscopic bodies in protoplanetary disks. Annu. Rev.
Astron. Astrophys. 46:21-56

\hfill
\vskip 0.5pt
\noindent
Blum J, Wurm G, Poppe T, Heim L-O. 1998. Aspects of laboratory dust aggregation with relevance to the
formation of planetesimals. Earth Moon Planets 80:285-309

\hfill
\vskip 0.5pt
\noindent
Boley AC. 2009. The two modes of gas-giant planet formation. Astrophys. J. 695:L53-57

\hfill
\vskip 0.5pt
\noindent
Boley AC, Hayfield T,Mayer L, Durisen RH. 2010. Clumps in the outer disk by disk instability: why they are
initially gas giants and the legacy of disruption. Icarus 207:509-16

\hfill
\vskip 0.5pt
\noindent
Bonfils X, Forveille T, Delfosse X, Udry S, Mayor M. 2005. The HARPS search for southern extra-solar
planets. VI. A Neptune-mass planet around the nearby M dwarf Gl 581. Astron. Astrophys. 443:L15-18

\hfill
\vskip 0.5pt
\noindent
Boss AP. 2000a. Formation of extrasolar giant planets: core accretion or disk instability? Earth Moon Planets
81:19-26

\hfill
\vskip 0.5pt
\noindent
Boss AP. 2000b. Possible rapid gas-giant planet formation in the solar nebula and other protoplanetary disks.
Astrophys. J. 536:L101-4

\hfill
\vskip 0.5pt
\noindent
Boss AP. 2003. Rapid formation of outer giant planets by disk instability. Astrophys. J. 599:577-81

\hfill
\vskip 0.5pt
\noindent
Boss AP. 2006. Rapid formation of super-Earths around M dwarf stars. Astrophys. J. 644:L79-82

\hfill
\vskip 0.5pt
\noindent
Boss AP,Wetherill GW, Haghighipour N. 2002. Rapid formation of ice giant planets. Icarus 156:291-95

\hfill
\vskip 0.5pt
\noindent
Brauer F, Dullemond CP, Henning T. 2008a. Coagulation, fragmentation and radial motion of solid particles
in protoplanetary disks. Astron. Astrophys. 480:859-77

\hfill
\vskip 0.5pt
\noindent
Brauer F, Henning T, Dullemond CP. 2008b. Planetesimal formation near the snow line in MRI-driven
turbulent protoplanetary disks. Astron. Astrophys. 487:L1-4

\hfill
\vskip 0.5pt
\noindent
Bromley BC, Kenyon SJ. 2006. Terrestrial planet formation. I. The transition from oligarchic growth to
chaotic growth. Astron. J. 131:1837-50

\hfill
\vskip 0.5pt
\noindent
Bromley BC, Kenyon SJ. 2011. A new hybrid N-body-coagulation code for the formation of gas giant planets.
Astrophys. J. 731:101

\hfill
\vskip 0.5pt
\noindent
Cai K, Pickett MK, Durisen RH, Milne AM. 2010. Giant planet formation by disk instability: a comparison
simulation with an improved radiative scheme. Astrophys. J. 716:L176-80

\hfill
\vskip 0.5pt
\noindent
Canup RM, Ward WR. 2002. Formation of the Galilean satellites: conditions of accretion. Astron. J. 124:3404-23

\hfill
\vskip 0.5pt
\noindent
Chambers JE. 2001. Making more terrestrial planets. Icarus 152:205-24

\hfill
\vskip 0.5pt
\noindent
Chambers JE. 2009. Planetary migration: What does it mean for planet formation? Annu. Rev. Earth Planet.
Sci. 37:321-44

\hfill
\vskip 0.5pt
\noindent
Chambers JE. 2010. Planetesimal formation by turbulent concentration. Icarus 208:505-17

\hfill
\vskip 0.5pt
\noindent
Chambers JE, Cassen P. 2002. The effect of surface density profile and giant planet eccentricities on planetary
accretion in the inner solar system. Meteorit. Planet. Sci. 37:1523-40

\hfill
\vskip 0.5pt
\noindent
Chambers JE, Wetherill GW. 1998. Making the terrestrial planets: N-body integrations of planetary embryos
in three dimensions. Icarus 136:304-27

\hfill
\vskip 0.5pt
\noindent
Chambers JE, Wetherill GW. 2001. Planets in the asteroid belt. Meteorit. Planet. Sci. 36:381–99
Charbonneau D, Berta ZK, Irwin J, Burke CJ, Nutzman P, et al. 2009. A super-Earth transiting a nearby
low-mass star. Nature 462:891-94

\hfill
\vskip 0.5pt
\noindent
Chen CH, Kamp I. 2004. Are giant planets forming around HR 4796A? Astrophys. J. 602:985-92

\hfill
\vskip 0.5pt
\noindent
Chiang E, Laughlin G. 2013. The minimum-mass extrasolar nebula: in-situ formation of close-in super-Earths.
MNRAS 413:3444-55

\hfill
\vskip 0.5pt
\noindent
Chiang E, Youdin AN. 2010. Forming planetesimals in solar and extrasolar nebulae. Annu. Rev. Earth Planet.
Sci. 38:493-522

\hfill
\vskip 0.5pt
\noindent
Clubb K, Fischer D, Howard A, Marcy G, Henry G. 2009. M2K: a search for planets orbiting early M and
late K dwarf stars. Bull. Am. Astron. Soc. 41:192

\hfill
\vskip 0.5pt
\noindent
Cuzzi JN, Hogan RC, Shariff K. 2008. Towards planetesimals: dense chondrule clumps in the protoplanetary
nebula. Astrophys. J. 687:1432-47

\hfill
\vskip 0.5pt
\noindent
Danby JMA. 1992. Fundamentals of Celestial Mechanics. Richmond, VA: Willmann-Bell. 2nd ed.

\hfill
\vskip 0.5pt
\noindent
D\'esert J-M, Bean J, Miller-Ricci Kempton E, Berta ZK, Charbonneau D, et al. 2011. Observational evidence
for a metal-rich atmosphere on the super-Earth GJ1214b. Astrophys. J. 731:L40

\hfill
\vskip 0.5pt
\noindent
Dominik C, Tielens AGGM. 1997. The physics of dust coagulation and the structure of dust aggregates in
space. Astrophys. J. 480:647-73

\hfill
\vskip 0.5pt
\noindent
Dressing CD, Charbonneau D. 2013. The occurrence rate of small planets around small stars. Astrophys. J. 767:95 

\hfill
\vskip 0.5pt
\noindent
Dullemond CP, Dominik C. 2005. Dust coagulation in protoplanetary disks: a rapid depletion of small grains.
Astron. Astrophys. 434:971-86

\hfill
\vskip 0.5pt
\noindent
Durisen RH, Boss AP, Mayer L, Nelson AF, Quinn T, Rice WKM. 2007. Gravitational instabilities in gaseous
protoplanetary disks and implications for giant planet formation. In Protostars and PlanetsV, ed. B Reipurth,
D Jewitt, K Keil, pp. 607-22. Tucson: Univ. Ariz. Press

\hfill
\vskip 0.5pt
\noindent
Fernandez JA, Ip W-H. 1984. Some dynamical aspects of the accretion of Uranus and Neptune-the exchange
of orbital angular momentum with planetesimals. Icarus 58:109-20

\hfill
\vskip 0.5pt
\noindent
Fogg MJ, Nelson RP. 2005. Oligarchic and giant impact growth of terrestrial planets in the presence of
gas-giant planet migration. Astron. Astrophys. 441:791-806

\hfill
\vskip 0.5pt
\noindent
Fogg MJ, Nelson RP. 2006. On the possibility of terrestrial planet formation in hot-Jupiter systems. Int. J.
Astrobiol. 5:199-209

\hfill
\vskip 0.5pt
\noindent
Fogg MJ, Nelson RP. 2007a. On the formation of terrestrial planets in hot-Jupiter systems. Astron. Astrophys.
461:1195-208

\hfill
\vskip 0.5pt
\noindent
Fogg MJ, NelsonRP. 2007b. The effect of type I migration on the formation of terrestrial planets in hot-Jupiter
systems. Astron. Astrophys. 472:1003-15

\hfill
\vskip 0.5pt
\noindent
Fogg MJ, Nelson RP. 2009. Terrestrial planet formation in low-eccentricity warm-Jupiter systems. Astron.
Astrophys. 498:575-89

\hfill
\vskip 0.5pt
\noindent
Forveille T, Bonfils X, Delfosse X, Alonso R, Udry S, et al. 2011. The HARPS search for southern extrasolar
planets. XXXII. Only 4 planets in GL 581 system. arXiv:1109.2505. http://arxiv.org/abs/1109.2505

\hfill
\vskip 0.5pt
\noindent
Fraine JD, Deming D, Gillon M, Jehin E, Demory B-O, et al. 2013. Spitzer transits of the super-Earth GJ
1214 b and implications for its atmosphere. Astrophys. J. 765:127

\hfill
\vskip 0.5pt
\noindent
Gibbons PG, Rice WKM, Mamatsashvili GR. 2012. Planetesimal formation in self-gravitating discs.
MNRAS 426:1444-54

\hfill
\vskip 0.5pt
\noindent
Goldreich P. 1965. An explanation of the frequent occurrence of commensurable mean motions in the Solar
System. MNRAS 130:159-81

\hfill
\vskip 0.5pt
\noindent
Goldreich P, Lithwick Y, Sari R. 2004. Final stages of planet formation. Astrophys. J. 614:497-507

\hfill
\vskip 0.5pt
\noindent
Gomes R, Levison HF, Tsiganis K, Morbidelli A. 2005. Origin of the cataclysmic Late Heavy Bombardment
period of the terrestrial planets. Nature 435:466-69

\hfill
\vskip 0.5pt
\noindent
Gomes RS. 1997. Dynamical effects of planetary migration on the primordial asteroid belt. Astron. J. 114:396-401

\hfill
\vskip 0.5pt
\noindent
Greenberg RJ. 1973. Evolution of satellite resonances by tidal dissipation. Astron. J. 78:338-46

\hfill
\vskip 0.5pt
\noindent
Greenberg RJ, Counselman CC, Shapiro II. 1972. Orbit-orbit resonance capture in the Solar System. Science
178:747-49

\hfill
\vskip 0.5pt
\noindent
Greenberg R, Hartmann WK, Chapman CR, Wacker JF. 1978. Planetesimals to planets-numerical simulations
of collisional evolution. Icarus 35:1-26

\hfill
\vskip 0.5pt
\noindent
Gressel O, Nelson RP, Turner NJ. 2012. Dead zones as safe havens for planetesimals: influence of disc mass
and external magnetic field. MNRAS 422:1140-59

\hfill
\vskip 0.5pt
\noindent
Guillot T. 2005. The interiors of giant planets: models and outstanding questions. Annu. Rev. Earth Planet.
Sci. 33:493-530

\hfill
\vskip 0.5pt
\noindent
G\"uttler C, Blum J, Zsom A, Ormel CW, Dullemond CP. 2009. The first phase of protoplanetary dust growth:
the bouncing barrier. Geochim. Cosmochim. Acta 73:A482

\hfill
\vskip 0.5pt
\noindent
Haghighipour N. 2005. Growth and sedimentation of dust particles in the vicinity of local pressure enhancements
in a solar nebula. MNRAS 362:1015-24

\hfill
\vskip 0.5pt
\noindent
Haghighipour N. 2011. Super-Earths: a new class of planetary bodies. Contemp. Phys. 52:403-38

\hfill
\vskip 0.5pt
\noindent
Haghighipour N, Boss AP. 2003a. On pressure gradients and rapid migration of solids in a nonuniform solar
nebula. Astrophys. J. 583:996-1003

\hfill
\vskip 0.5pt
\noindent
Haghighipour N, Boss AP. 2003b. On gas drag–induced rapid migration of solids in a nonuniform solar
nebula. Astrophys. J. 598:1301-11

\hfill
\vskip 0.5pt
\noindent
Haghighipour N, Jewitt D. 2008. A region void of irregular satellites around Jupiter. Astron. J. 136:909-18

\hfill
\vskip 0.5pt
\noindent
Haghighipour N, Rastegar S. 2011. Implications of the TTV-detection of close-in terrestrial planets around
M stars for their origin and dynamical evolution. Proc. Haute-Provence Obs. Colloq., St. Michel l'Observatoire,
France, Aug. 23–27, 2010, ed. F Bouchy, RF Diaz, C Moutou, 11:04004. Les Ulis, France: EPJ Web of
Conferences. http://dx.doi.org/10.1051/epjconf/20101104004

\hfill
\vskip 0.5pt
\noindent
Haghighipour N, Vogt SS, Butler RP, Rivera EJ, Laughlin G, et al. 2010. The Lick-Carnegie Exoplanet
Survey: a Saturn-mass planet in the habitable zone of the nearby M4V star HIP 57050. Astrophys. J.
715:271-76

\hfill
\vskip 0.5pt
\noindent
Hahn JM, Malhotra R. 2005. Neptune’s migration into a stirred-up Kuiper belt: a detailed comparison of
simulations to observations. Astron. J. 130:2392-414

\hfill
\vskip 0.5pt
\noindent
Haisch KE Jr, Lada EA, Lada CJ. 2001. Disk frequencies and lifetimes in young clusters. Astrophys. J. Lett.
553:L153-56

\hfill
\vskip 0.5pt
\noindent
Hansen BMS. 2009. Formation of the terrestrial planets from a narrow annulus. Astrophys. J. 703:1131-40

\hfill
\vskip 0.5pt
\noindent
Hatzes AP, Dvorak R, Wuchterl G, Guterman P, Hartmann M, et al. 2010. An investigation into the radial
velocity variations of CoRoT-7. Astron. Astrophys. 520:93-108

\hfill
\vskip 0.5pt
\noindent
Hatzes AP, Fridlund M, Nachmani G, Mazeh T, Valencia D, et al. 2011. The mass of CoRoT-7 b. Astrophys.
J. 743:75

\hfill
\vskip 0.5pt
\noindent
Hayashi C. 1981. Structure of the solar nebula, growth and decay of magnetic fields and effects of magnetic
and turbulent viscosities on the nebula. Prog. Theor. Phys. 70:S35-53

\hfill
\vskip 0.5pt
\noindent
Heng K, Vogt SS. 2011. Gliese 581g as a scaled-up version of Earth: atmospheric circulation simulations.
MNRAS 415:2145-57

\hfill
\vskip 0.5pt
\noindent
Hubickyj O, Bodenheimer P, Lissauer JJ. 2005. Accretion of the gaseous envelope of Jupiter around a 5-10
Earth-mass core. Icarus 179:415-31

\hfill
\vskip 0.5pt
\noindent
Ida S, Makino J. 1993. Scattering of planetesimals by a protoplanet: slowing down of runaway growth. Icarus
106:210-27

\hfill
\vskip 0.5pt
\noindent
Irwin J, Charbonneau D, Nutzman P, Falco E. 2009a. The MEarth Project: searching for transiting habitable
super-Earth planets around nearby M-dwarfs. Proc. 15th Cambridge Workshop Cool Stars Stellar Syst. Sun,
St. Andrews, Scotl., July 21-25, 2008. AIP Conf. Proc. 1094:445-48. Melville, NY: AIP

\hfill
\vskip 0.5pt
\noindent
Irwin J, Charbonneau D, Nutzman P, Falco E. 2009b. The MEarth Project: searching for transiting habitable
super-Earths around nearby M-dwarfs. Proc. Int. Astron. Union Symp., Boston, May 19-23, 2008. 253:37-43.
Cambridge, UK: Cambridge Univ. Press

\hfill
\vskip 0.5pt
\noindent
Johansen A, Brauer F, Dullemond C, Klahr H, Henning T. 2008. A coagulation-fragmentation model for the
turbulent growth and destruction of preplanetesimals. Astron. Astrophys. 486:597-611

\hfill
\vskip 0.5pt
\noindent
Johansen A, Klahr H,Henning T. 2006. Gravoturbulent formation of planetesimals. Astrophys. J. 636:1121-34

\hfill
\vskip 0.5pt
\noindent
Johansen A, Oishi JS, Mac Low M-M, Klahr H, Henning T, Youdin A. 2007. Rapid planetesimal formation
in turbulent circumstellar disks. Nature 448:1022-25

\hfill
\vskip 0.5pt
\noindent
Johnson JA, Gazak JZ, Apps K, Muirhead PS, Crepp JR, et al. 2012. Characterizing the cool KOIs. II. The
M dwarf KOI-254 and its hot Jupiter. Astron. J. 143:111

\hfill
\vskip 0.5pt
\noindent
Kato MT, Fujimoto M, Ida S. 2012. Planetesimal formation at the boundary between steady super/sub-Keplerian 
flow created by inhomogeneous growth of magnetorotational instability. Astrophys. J. 747:11-20

\hfill
\vskip 0.5pt
\noindent
Kennedy GM, Kenyon SJ. 2008a. Planet formation around stars of various masses: the snow line and the
frequency of giant planets. Astrophys. J. 673:502-12

\hfill
\vskip 0.5pt
\noindent
Kennedy GM, Kenyon SJ. 2008b. Planet formation around stars of various masses: hot super-Earths. Astrophys.
J. 682:1264-76

\hfill
\vskip 0.5pt
\noindent
Kennedy GM, Kenyon SJ, Bromley BC. 2006. Planet formation around low-mass stars: the moving snow line
and super-Earths. Astrophys. J. 650:L139-42

\hfill
\vskip 0.5pt
\noindent
Kennedy GM, Kenyon SJ, Bromley BC. 2007. Planet formation around M-dwarfs: the moving snow line and
super-Earths. Astrophys. Space Sci. 311:9-13

\hfill
\vskip 0.5pt
\noindent
Kenyon SJ, Bromley BC. 2009. Rapid formation of icy super-Earths and the cores of gas-giant planets.
Astrophys. J. 690:L140-43

\hfill
\vskip 0.5pt
\noindent
Klahr HH, Henning T. 1997. Particle-trapping eddies in protoplanetary accretion disks. Icarus 128:213-29

\hfill
\vskip 0.5pt
\noindent
Kokubo E, Ida S. 1995. Orbital evolution of protoplanets embedded in a swarm of planetesimals. Icarus
114:247-57

\hfill
\vskip 0.5pt
\noindent
Kokubo E, Ida S. 1996. On runaway growth of planetesimals. Icarus 123:180-91

\hfill
\vskip 0.5pt
\noindent
Kokubo E, Ida S. 1998. Orbital oligarchic growth of protoplanets. Icarus 131:171–78

\hfill
\vskip 0.5pt
\noindent
Kokubo E, Ida S. 2000. Formation of protoplanets from planetesimals in the solar nebula. Icarus 143:15-27

\hfill
\vskip 0.5pt
\noindent
Kokubo E, Ida S. 2007. Formation of terrestrial planets from protoplanets. II. Statistics of planetary spin.
Astrophys. J. 671:2082-90

\hfill
\vskip 0.5pt
\noindent
Kokubo E, Kominami J, Ida S. 2006. Formation of terrestrial planets from protoplanets. I. Statistics of basic
dynamical properties. Astrophys. J. 642:1131-39

\hfill
\vskip 0.5pt
\noindent
Krause M, Blum J. 2004. Growth and form of planetary seedlings: results from a sounding rocket microgravity
aggregation experiment. Phys. Rev. Lett. 93:021103

\hfill
\vskip 0.5pt
\noindent
Kretke KA, Lin DNC. 2007. Grain retention and formation of planetesimals near the snow line inMRI-driven
turbulent protoplanetary disks. Astrophys. J. 664:L55-58

\hfill
\vskip 0.5pt
\noindent
Lada CJ, Lada EA. 2003. Embedded clusters in molecular clouds. Annu. Rev. Astron. Astrophys. 41:57-115

\hfill
\vskip 0.5pt
\noindent
Laughlin G, Bodenheimer P, Adams FC. 2004. The core-accretion model predicts few Jovian-mass planets
orbiting red dwarfs. Astrophys. J. 612:L73-76

\hfill
\vskip 0.5pt
\noindent
L\'eger A, Rouan D, Schneider J, Barge P, Fridlund M, et al. 2009. Transiting exoplanets from the CoRoT space
mission. VIII. CoRoT-7b: the first super-Earth with measured radius. Astron. Astrophys. 506:287-302

\hfill
\vskip 0.5pt
\noindent
Levison HF, Agnor C. 2003. The role of giant planets in terrestrial planet formation. Astron. J. 125:2692-713

\hfill
\vskip 0.5pt
\noindent
Lissauer JJ, Hubickyj O, D'Angelo G, Bodenheimer P. 2009. Models of Jupiter’s growth incorporating thermal
and hydrodynamic constraints. Icarus 199:338-50

\hfill
\vskip 0.5pt
\noindent
Lyra W, Johansen A, Klahr H, Piskunov N. 2009a. Standing on the shoulders of giants. Trojan earths and
vortex trapping in low mass self-gravitating protoplanetary disks of gas and solids. Astron. Astrophys.
493:1125-39

\hfill
\vskip 0.5pt
\noindent
Lyra W, Johansen A, Zsom A, Klahr H, Piskunov N. 2009b. Planet formation bursts at the borders of the
dead zone in 2D numerical simulations of circumstellar disks. Astron. Astrophys. 497:869-88

\hfill
\vskip 0.5pt
\noindent
Maercker M, Burton MG, Right WCM. 2006. L-band (3.5 μm) IR-excess in massive star formation. II. RCW
57/NGC 3576. Astron. Astrophys. 450:253-63

\hfill
\vskip 0.5pt
\noindent
Malhotra R. 1993. The origin of Pluto’s peculiar orbit. Nature 365:819-21

\hfill
\vskip 0.5pt
\noindent
Malhotra R. 1995. The origin of Pluto’s orbit: implications for the Solar System beyond Neptune. Astron. J.
110:420-29

\hfill
\vskip 0.5pt
\noindent
Malhotra R. 1996. The phase space structure near Neptune resonances in the Kuiper belt. Astron. J. 111:504-16

\hfill
\vskip 0.5pt
\noindent
Mandell AM, Raymond SN, Sigurdsson S. 2007. Formation of Earth-like planets during and after giant planet
migration. Astrophys. J. 660:823-44

\hfill
\vskip 0.5pt
\noindent
Mandell AM, Sigurdsson S. 2003. Survival of terrestrial planets in the presence of giant planet migration.
Astrophys. J. 599:L111-14

\hfill
\vskip 0.5pt
\noindent
Mass\'et F, Snellgrove M. 2001. Reversing type II migration: resonance trapping of a lighter giant protoplanet.
MNRAS 320:L55-59

\hfill
\vskip 0.5pt
\noindent
Mayer L, Lufkin G, Quinn T, Wadsley J. 2007. Fragmentation of gravitationally unstable gaseous protoplanetary
disks with radiative transfer. Astrophys. J. 661:L77-80

\hfill
\vskip 0.5pt
\noindent
Mayer L, Quinn T, Wadsley J, Stadel J. 2002. Formation of giant planets by fragmentation of protoplanetary
disks. Science 298:1756-59

\hfill
\vskip 0.5pt
\noindent
Mayer L, Quinn T, Wadsley J, Stadel J. 2003. Simulations of unstable gaseous disks and the origin of giant
planets. In ASP Conference Series. Vol. 294: Scientific Frontiers in Research on Extrasolar Planets, ed. D
Deming, S Seager, pp. 281–86. San Francisco: Astron. Soc. Pac.

\hfill
\vskip 0.5pt
\noindent
Mayer L, Quinn T, Wadsley J, Stadel J. 2004. The evolution of gravitationally unstable protoplanetary disks:
fragmentation and possible giant planet formation. Astrophys. J. 609:1045-64

\hfill
\vskip 0.5pt
\noindent
MayorM, Bonfils X, Forveille T, Delfosse X, Udry S, et al. 2009. The HARPS search for southern extrasolar
planets. XVIII. An Earth-mass planet in the GJ 581 planetary system. Astron. Astrophys. 507:487-94

\hfill
\vskip 0.5pt
\noindent
Mayor M, Queloz D. 1995. A Jupiter-mass companion to a solar-type star. Nature 378:355-59

\hfill
\vskip 0.5pt
\noindent
McNeil DS, Nelson RP. 2010. On the formation of hot Neptunes and super-Earths. MNRAS 401:1691-708

\hfill
\vskip 0.5pt
\noindent
Menou K. 2012. Atmospheric circulation and composition of GJ 1214 b. Astrophys. J. 744:L16

\hfill
\vskip 0.5pt
\noindent
Michael S, Durisen RH, Boley AC. 2011. Migration of gas giant planets in gravitationally unstable disks.
Astrophys. J. 737:L42

\hfill
\vskip 0.5pt
\noindent
Militzer B, Hubbard WB, Vorberger J, Tamblyn I, Bonev SA. 2008. A massive core in Jupiter predicted from
first-principles simulations. Astrophys. J. 688:L45-48

\hfill
\vskip 0.5pt
\noindent
Miller-Ricci E, Fortney JJ. 2010. The nature of the atmosphere of the transiting super-Earth GJ 1214 b.
Astrophys. J. Lett. 716:L74-79

\hfill
\vskip 0.5pt
\noindent
Miller-Ricci E, Seager S, Sasselov D. 2009. The atmospheric signatures of super-Earths: how to distinguish
between hydrogen-rich and hydrogen-poor atmospheres. Astrophys. J. 690:1056-67

\hfill
\vskip 0.5pt
\noindent
Minton DA, Malhotra R. 2009. A record of planet migration in the main asteroid belt. Nature 457:1109-11

\hfill
\vskip 0.5pt
\noindent
Minton DA, Malhotra R. 2011. Secular resonance sweeping of the main asteroid belt during planet migration.
Astrophys. J. 732:53

\hfill
\vskip 0.5pt
\noindent
Mizuno H, Markiewicz WJ, V\"olk HJ. 1988. Grain growth in turbulent protoplanetary accretion disks. Astron.
Astrophys. 195:183-92

\hfill
\vskip 0.5pt
\noindent
Morbidelli A, Chambers J, Lunine JL, Petit JM, Robert F, et al. 2000. Source regions and timescales for the
delivery of water to Earth. Meteorit. Planet. Sci. 35:1309-20

\hfill
\vskip 0.5pt
\noindent
Morbidelli A, Levison HF, Tsiganis K, Gomes R. 2005. Chaotic capture of Jupiter’s Trojan asteroids in the
early Solar System. Nature 435:462-65

\hfill
\vskip 0.5pt
\noindent
Morbidelli A, Lunine JI, O'Brien DP, Raymond SN, Walsh KJ. 2012. Building terrestrial planets. Annu. Rev.
Earth Planet. Sci. 40:251-75

\hfill
\vskip 0.5pt
\noindent
Movshovitz N, Bodenheimer P, Podolak M, Lissauer JJ. 2010. Formation of Jupiter using opacities based on
detailed grain physics. Icarus 209:616-24

\hfill
\vskip 0.5pt
\noindent
Movshovitz N, Podolak M. 2008. The opacity of grains in protoplanetary atmospheres. Icarus 194:368-78

\hfill
\vskip 0.5pt
\noindent
Murray CD, Dermott SF. 1999. Solar System Dynamics. Cambridge, UK: Cambridge Univ. Press

\hfill
\vskip 0.5pt
\noindent
Nakagawa Y, Nakazawa K, Hayashi C. 1981. Growth and sedimentation of dust grains in the primordial solar
nebula. Icarus 45:517-28

\hfill
\vskip 0.5pt
\noindent
Nakagawa Y, Sekiya M, Hayashi C. 1986. Settling and growth of dust particles in a laminar phase of a low-mass
solar nebula. Icarus 67:375-90

\hfill
\vskip 0.5pt
\noindent
Nelson RP, Papaloizou JCB, Mass\'et F, Kley W. 2001. The migration and growth of protoplanets in protostellar
discs. MNRAS 318:18-36

\hfill
\vskip 0.5pt
\noindent
Nutzman P, Charbonneau D. 2008. Design considerations for a ground-based transit search for habitable
planets orbiting M dwarfs. Publ. Astron. Soc. Pac. 120:317-27

\hfill
\vskip 0.5pt
\noindent
O'Brien DP, Morbidelli A, Levison HF. 2006. Terrestrial planet formation with strong dynamical friction.
Icarus 184:39-58

\hfill
\vskip 0.5pt
\noindent
Okuzumi S, Tanaka H, Kobayashi H, Wada K. 2012. Rapid coagulation of porous dust aggregates outside
the snow line: a pathway to successful icy planetesimal formation. Astrophys. J. 752:106-23

\hfill
\vskip 0.5pt
\noindent
O'Neill C, Lenardic A. 2007. Geological consequences of super-sized Earths. Geophys. Res. Lett. 34:L19204

\hfill
\vskip 0.5pt
\noindent
Ormel CW, Cuzzi JN. 2007. Closed-form expressions for particle relative velocities induced by turbulence.
Astrophys. Astron. 466:413-20

\hfill
\vskip 0.5pt
\noindent
Pan L, Padoan P, Scalo J, Kritsuk AG, Norman ML. 2011. Turbulent clustering of protoplanetary dust and
planetesimal formation. Astrophys. J. 740:6

\hfill
\vskip 0.5pt
\noindent
Papaloizou JCB, Terquem C. 2006. Planet formation and migration. Rep. Prog. Phys. 69:119-80

\hfill
\vskip 0.5pt
\noindent
Peale SJ, Lee MH. 2002. A primordial origin of the Laplace relation among the Galilean satellites. Science
298:593-97

\hfill
\vskip 0.5pt
\noindent
Podolak M. 2003. The contribution of small grains to the opacity of protoplanetary atmospheres. Icarus
165:428-37

\hfill
\vskip 0.5pt
\noindent
Pollack JB, Hubickyj O, Bodenheimer P, Lissauer JJ, Podolak M, Greenzweig Y. 1996. Formation of the giant
planets by concurrent accretion of solids and gas. Icarus 124:62-85

\hfill
\vskip 0.5pt
\noindent
Queloz D, Bouchy F, Moutou C, Hatzes A, H\'ebrard G, et al. 2009. The CoRoT-7 planetary system: two
orbiting super-Earths. Astron. Astrophys. 506:303-19

\hfill
\vskip 0.5pt
\noindent
Raymond SN, Barnes R, Mandell AM. 2008. Observable consequences of planet formation models in systems
with close-in terrestrial planets. MNRAS 384:663-74

\hfill
\vskip 0.5pt
\noindent
Raymond SN, Mandell AM, Sigurdsson S. 2006a. Exotic earths: forming habitable worlds with giant planet
migration. Science 313:1413-16

\hfill
\vskip 0.5pt
\noindent
Raymond SN, O'Brien DP, Morbidelli A, Kaib NA. 2009. Building the terrestrial planets: constrained accretion
in the inner Solar System. Icarus 203:644-62

\hfill
\vskip 0.5pt
\noindent
Raymond SN, Quinn T, Lunine JI. 2004. Making other earths: dynamical simulations of terrestrial planet
formation and water delivery. Icarus 168:1-17

\hfill
\vskip 0.5pt
\noindent
Raymond SN, Quinn T, Lunine JI. 2005a. Terrestrial planet formation in disks with varying surface density
profiles. Astrophys. J. 632:670-76

\hfill
\vskip 0.5pt
\noindent
Raymond SN, Quinn T, Lunine JI. 2005b. The formation and habitability of terrestrial planets in the presence
of close-in giant planets. Icarus 177:256-63

\hfill
\vskip 0.5pt
\noindent
Raymond SN, Quinn T, Lunine JI. 2006b. High-resolution simulations of the final assembly of Earth-like
planets 1: terrestrial accretion and dynamics. Icarus 183:265-82

\hfill
\vskip 0.5pt
\noindent
Raymond SN, Quinn T, Lunine JI. 2007. High resolution simulations of the final assembly of Earth-like
planets 2: water delivery and planetary habitability. Astrobiol. J. 7:66-84

\hfill
\vskip 0.5pt
\noindent
Rivera EJ, Laughlin G, Butler RP, Vogt SS, Haghighipour N, Meschiari S. 2010. TheLick-CarnegieExoplanet
Survey: a Uranus-mass fourth planet for GJ 876 in an extrasolar Laplace configuration. Astrophys. J.
719:890-99

\hfill
\vskip 0.5pt
\noindent
Rivera EJ, Lissauer JJ, Butler RP, Marcy GW, Vogt SS, et al. 2005. A $\sim 7.5 {M_\oplus}$ planet orbiting the nearby
star, GJ 876. Astrophys. J. 634:625-40

\hfill
\vskip 0.5pt
\noindent
Rogers LA, Bodenheimer P, Lissauer J, Seager S. 2011. Formation and structure of low-density exo-Neptunes.
Astrophys. J. 738:59

\hfill
\vskip 0.5pt
\noindent
Rogers LA, Seager S. 2010a. A framework for quantifying the degeneracies of exoplanet interior compositions.
Astrophys. J. 712:974-91

\hfill
\vskip 0.5pt
\noindent
Rogers LA, Seager S. 2010b. Three possible origins for the gas layer on GJ 1214b. Astrophys. J. 716:1208-16

\hfill
\vskip 0.5pt
\noindent
Roy AE. 1982. Orbital Motion. Bristol, UK: Adam Hilger

\hfill
\vskip 0.5pt
\noindent
Safronov VS. 1969. Evolution of Protoplanetary Cloud and Formation of the Earth and Planets. Moscow: Nauka

\hfill
\vskip 0.5pt
\noindent
Schlichting E, Warren PH, Yin Q-Z. 2012. The last stages of terrestrial planet formation: dynamical friction
and the late veneer. Astrophys. J. 752:8-15

\hfill
\vskip 0.5pt
\noindent
Seager S, Deming D. 2009. On the method to infer an atmosphere on a tidally locked super Earth exoplanet
and upper limits to GJ 876d. Astrophys. J. 703:1884-89

\hfill
\vskip 0.5pt
\noindent
Shariff K, Cuzzi JN. 2011. Gravitational instability of solids assisted by gas drag: slowing by turbulent mass
diffusivity. Astrophys. J. 738:73

\hfill
\vskip 0.5pt
\noindent
Shi J-M, Chiang E. 2013. From dust to planetesimals: criteria for gravitational instability of small particles in
gas. Astrophys. J. 764:20

\hfill
\vskip 0.5pt
\noindent
Smoluchowski MV. 1916. Drei Vortrage uber diffusion, brownsche bewegung und koagulation von Kolloidteilchen.
Physik. Zeit. 17:557-85

\hfill
\vskip 0.5pt
\noindent
Sotin C, Schubert G. 2009. Mantle convection and plate tectonics on Earth-like exoplanets. Presented at Am.
Geophys. Union Fall Meet., Dec. 14-18, San Francisco (Abstr. P42B-02)

\hfill
\vskip 0.5pt
\noindent
Strom SE, Edwards S, Skrutskie MF. 1993. Evolutionary time scales for circumstellar disks associated with
intermediate- and solar-type stars. In Protostars and Planets III, ed. EH Levy, JI Lunine, pp. 837-66.
Tucson: Univ. Ariz. Press

\hfill
\vskip 0.5pt
\noindent
Supulver KD, Lin DNC. 2000. Growth and sedimentation of dust grains in the primordial solar nebula. Icarus
146:525-40

\hfill
\vskip 0.5pt
\noindent
Swift J, Johnson JA, Morton TD, Crepp JR, Montet BT, et al. 2013. Characterizing the cool KOIs. IV.Kepler-32 
as a prototype for the formation of compact planetary systems throughout the galaxy. Astrophys. J. 764:105

\hfill
\vskip 0.5pt
\noindent
Tackley PJ, van Heck H. 2009. Mantle convection, stagnant lids and plate tectonics on super-Earths. Geochim.
Cosmochim. Acta 73:A1303

\hfill
\vskip 0.5pt
\noindent
Terquem C, Papaloizou JCB. 2007. Migration and the formation of systems of hot super-Earths and Neptunes.
Astrophys. J. 654:1110-20

\hfill
\vskip 0.5pt
\noindent
Torres K, Winter OC, Izidoro A, Haghighipour N. 2013. A compound model for the origin of Earth’s water.
Astrophys. J. 767:54

\hfill
\vskip 0.5pt
\noindent
Tsiganis K, Gomes R, Morbidelli A, Levison HF. 2005. Origin of the orbital architecture of the giant planets
of the Solar System. Nature 435:459-61

\hfill
\vskip 0.5pt
\noindent
Udry S, Bonfils X, Delfosse X, Forveille T, Mayor M, et al. 2007. The HARPS search for southern extra-solar
planets. XI. Super-Earths (5 and 8 $M_\oplus$) in a 3-planet system. Astron. Astrophys. 469:L43-47

\hfill
\vskip 0.5pt
\noindent
Valencia D, Ikoma M, Guillot T, Nettelmann N. 2010. Composition and fate of short-period super-Earths.
The case of CoRoT-7b. Astron. Astrophys. 516:A20

\hfill
\vskip 0.5pt
\noindent
Valencia D, O'Conell RJ. 2009. Convection scaling and subduction on Earth and super-Earths. Earth Planet.
Sci. Lett. 286:492-502

\hfill
\vskip 0.5pt
\noindent
Valencia D, O'Connell RJ, Sasselov DD. 2006. Internal structure of massive terrestrial planets. Icarus 181:545-54

\hfill
\vskip 0.5pt
\noindent
Valencia D, O'Connell RJ, Sasselov DD. 2007a. Inevitability of plate tectonics on super-Earths. Astrophys. J.
670:L45-48

\hfill
\vskip 0.5pt
\noindent
Valencia D, Sasselov DD, O'Connell RJ. 2007b. Detailed models of super-Earths: How well can we infer bulk
properties? Astrophys. J. 665:1413-20

\hfill
\vskip 0.5pt
\noindent
Valencia D, Sasselov DD, O'Connell RJ. 2007c.Radius and structure models of the first terrestrial super-Earth
planets. Astrophys. J. 656:545-51

\hfill
\vskip 0.5pt
\noindent
Vogt SS, Butler RO, Haghighipour N. 2012. GJ 581 update: additional evidence for a super-Earth in the
habitable zone. Astron. Nachr. 333:561-75

\hfill
\vskip 0.5pt
\noindent
Vogt SS, Butler RP, Rivera EJ, Haghighipour N, Henry GW, Williamson MH. 2010. The Lick-Carnegie
Exoplanet Survey: A $3.1 {M_\oplus}$ planet in the habitable zone of the nearby M3V star Gliese 581. Astrophys.
J. 723:954-65

\hfill
\vskip 0.5pt
\noindent
V\"olk HJ, Jones FC, Morfill GE, Roeser S. 1980. Collisions between grains in a turbulent gas. Astron. Astrophys.
85:316-25

\hfill
\vskip 0.5pt
\noindent
Wada K, Tanaka H, Suyama T, Kimura H, Yamamoto T. 2007. Numerical simulation of dust aggregate
collisions. I. Compression and disruption of two-dimensional aggregates. Astrophys. J. 661:320-33

\hfill
\vskip 0.5pt
\noindent
Walsh KJ, Morbidelli A, Raymond SN, O'Brien DP, Mandell AM. 2011. A low mass for Mars from Jupiter’s
early gas-driven migration. Nature 475:206-9

\hfill
\vskip 0.5pt
\noindent
Weidenschilling SJ. 1977. Aerodynamics of solid bodies in the solar nebula. MNRAS 180:57-70

\hfill
\vskip 0.5pt
\noindent
Weidenschilling SJ. 1980. Growth and sedimentation of dust grains in the primordial solar nebula. Icarus
44:172-89

\hfill
\vskip 0.5pt
\noindent
Weidenschilling SJ. 2010. Particles in the nebular midplane: collective effects and relative velocities. Meteorit.
Planet. Sci. 45:276-88

\hfill
\vskip 0.5pt
\noindent
Weidenschilling SJ, Spaute D, Davis DR, Marzari F, Ohtsuki K. 1997. Accretional evolution of a planetesimal
swarm. Icarus 128:429-55

\hfill
\vskip 0.5pt
\noindent
Wetherill GW. 1990a. Comparison of analytical and physical modeling of planetesimal accumulation. Icarus
88:336-54

\hfill
\vskip 0.5pt
\noindent
Wetherill GW. 1990b. Formation of earth. Annu. Rev. Earth Planet. Sci. 18:205-56

\hfill
\vskip 0.5pt
\noindent
Wetherill GW. 1994. Provenance of the terrestrial planets. Geochim. Cosmochim. Acta 58:4513-20

\hfill
\vskip 0.5pt
\noindent
Wetherill GW. 1996. The formation and habitability of extra-solar planets. Icarus 119:219-38

\hfill
\vskip 0.5pt
\noindent
Wetherill GW, Stewart GR. 1989. Accumulation of a swarm of small planetesimals. Icarus 77:330-57

\hfill
\vskip 0.5pt
\noindent
Wetherill GW, Stewart GR. 1993. Formation of planetary embryos-effects of fragmentation, low relative
velocity, and independent variation of eccentricity and inclination. Icarus 106:190-209

\hfill
\vskip 0.5pt
\noindent
Whipple FL. 1972. On certain aerodynamic processes for asteroid and comets. Proc. Nobel Symp., 21st,
Saltsj\"obaden, Sweden, Sept. 6-10, 1971, ed. A Elvius, pp. 211-32. New York: Wiley

\hfill
\vskip 0.5pt
\noindent
Wolszczan A, Frail DA. 1992. A planetary system around the millisecond pulsar PSR 1257+12. Nature 355:145-47

\hfill
\vskip 0.5pt
\noindent
Wurm G, Blum J. 1998. Experiments on preplanetary dust aggregation. Icarus 132:125-36

\hfill
\vskip 0.5pt
\noindent
Zhou J-L, Aarseth SJ, Lin DNC, Nagasawa M. 2005. Origin and ubiquity of short-period Earth-like planets:
evidence for the sequential accretion theory of planet formation. Astrophys. J. 631:L85-88

\hfill
\vskip 0.5pt
\noindent
Zsom A, Ormel CW, G\"uttler C, Blum J, Dullemond CP. 2010. The first phase of protoplanetary dust growth:
the bouncing barrier. Astron. Astrophys. 513:A57


\clearpage

\begin{table}
\caption{Currently known extrasolar planets with masses up to 10 Earth-masses. 
The quantities $M, P, a$ and $e$ represent the mass (in terms of Earth's mass $M_\oplus$), 
orbital period, semimajor axis, and orbital eccentricity of the planet. The mass 
of the central star is shown by $M_*$ and is given in the units of solar-masses $(M_\odot)$.}
{\begin{tabular}{@{}lllllll}
\toprule 
Planet & $M (M_\oplus$) & $P$ (day) & $a$ (AU)& 
$\>e$ & Stellar Type &  ${M_*} (M_\odot)$ \\
\colrule 
KOI-55 c &$\>\>$ 	0.6678 &$\>\>$ 		0.34289 &$\>\>$	    0.0076 &	 -   &$\>\>$    0.496 &$\>\>$	  sdB	 \\	
Kepler-42 d &$\>\>$ 	0.954 &$\>\>$ 		1.856169 &$\>\>$    0.0154 &	 -   &$\>\>$	0.13 &$\>\>$	   -	 \\	
Kepler-42 c &$\>\>$ 	1.908 &$\>\>$ 		0.45328509 &$\>\>$  0.006  &	 -   &$\>\>$	0.13 &$\>\>$	   -	 \\	
Gl 581 e &$\>\>$ 	1.9398 &$\>\>$   	3.14945 &$\>\>$	    0.028  &	0.32 &$\>\>$	0.31 &$\>\>$	  M2.5V	 \\	
Kepler-11 f &$\>\>$ 	2.301366 &$\>\>$ 	46.68876 &$\>\>$    0.25   &	0    &$\>\>$	0.95 &$\>\>$	  G	 \\	
HD 20794 c &$\>\>$ 	2.4168 &$\>\>$ 		40.114 &$\>\>$	    0.2036 &	0    &$\>\>$	0.7 &$\>\>$	  G8V	 \\	
HD 20794 b &$\>\>$ 	2.703 &$\>\>$ 		18.315 &$\>\>$	    0.1207 &	0    &$\>\>$	0.7 &$\>\>$	  G8V	 \\	
HD 215152 b &$\>\>$ 	2.7666 &$\>\>$ 		7.2825 &$\>\>$	    0.0652 &	0.34 &$\>\>$	-   &$\>\>$	  K0	 \\	
Kepler-42 b &$\>\>$ 	2.862 &$\>\>$ 		1.2137672 &$\>\>$   0.0116 &	-    &$\>\>$	0.13 &$\>\>$	  -	 \\	
HD 215152 c &$\>\>$ 	3.0846 &$\>\>$ 		10.866 &$\>\>$	    0.0852 &	0.38 &$\>\>$	-    &$\>\>$	  K0	 \\	
Kepler-20 e &$\>\>$ 	3.0846 &$\>\>$ 		6.098493 &$\>\>$    0.0507 &	     &$\>\>$	0.912 &$\>\>$	  G8	 \\	
MOA-2007-BLG  &$\>\>$ 3.18 &$\>\>$ - &$\>\>$	    0.66   &	-    &$\>\>$	0.06 &$\>\>$	  M	 \\	
$\qquad\quad$-192-L b \\
Kepler-32 b &$\>\>$     3.4    &$\>\>$          5.90     &$\>\>$    0.0519 &    -    &$\>\>$     0.54&$\>\>$      M1V    \\
HD 85512 b &$\>\>$ 	3.498 &$\>\>$ 		58.43 &$\>\>$	    0.26   &	0.11 &$\>\>$	0.69 &$\>\>$   	  K5V	 \\	
HD 39194 b &$\>\>$ 	3.7206 &$\>\>$ 		5.6363 &$\>\>$	    0.0519 &	0.2  &$\>\>$	-    &$\>\>$	  K0V	 \\	
Kepler-32 c &$\>\>$     3.8    &$\>\>$          8.75   &$\>\>$      0.067  &    -    &$\>\>$    0.54 &$\>\>$      M1V    \\
PSR 1257 +12 d &$\>\>$ 	3.816 &$\>\>$ 		98.2114 &$\>\>$	    0.46   &	0.025&$\>\>$	-    &$\>\>$	  -	 \\	
PSR 1257 +12 c &$\>\>$ 	4.134 &$\>\>$ 		66.5419 &$\>\>$	    0.36   &	0.018&$\>\>$	-    &$\>\>$	  -	 \\	
HD 156668 b &$\>\>$ 	4.1658 &$\>\>$ 		4.646 &$\>\>$	    0.05   &	0    &$\>\>$	0.772 &$\>\>$	  K3V	 \\	
HD 40307 b &$\>\>$ 	4.1976 &$\>\>$ 		4.3115 &$\>\>$	    0.047  &	0    &$\>\>$	0.77 &$\>\>$	  K2.5V	 \\	
GJ 667C c &$\>\>$ 	4.2612 &$\>\>$ 		28.13 &$\>\>$	    0.1251 &	0.34 &$\>\>$	0.33 &$\>\>$	  M1.5V	 \\	
Kepler-11 b &$\>\>$ 	4.30254 &$\>\>$ 	10.30375 &$\>\>$    0.091  &	0    &$\>\>$	0.95 &$\>\>$	  G	 \\	
KOI-55 b &$\>\>$ 	4.452 &$\>\>$ 		0.2401 &$\>\>$	    0.006  &	-    &$\>\>$	0.496 &$\>\>$	  sdB	 \\	
Kepler-10 b &$\>\>$ 	4.5474 &$\>\>$ 		0.837495 &$\>\>$    0.01684 & 	0    &$\>\>$	0.895 &$\>\>$	  G	 \\	
HD 20794 d &$\>\>$ 	4.77 &$\>\>$ 		90.309 &$\>\>$	    0.3499 &	0    &$\>\>$	0.7 &$\>\>$	  G8V	 \\	
CoRoT-7 b &$\>\>$ 	4.8018 &$\>\>$ 		0.853585 &$\>\>$    0.0172 &	0    &$\>\>$	0.93 &$\>\>$	  K0V	 \\	
61 Vir b &$\>\>$ 	5.088 &$\>\>$ 		4.215 &$\>\>$	    0.050201 &	0.12 &$\>\>$	0.95 &$\>\>$	  G5V	 \\	
HD 39194 d &$\>\>$ 	5.1516 &$\>\>$ 		33.941 &$\>\>$	    0.172  &	0.2  &$\>\>$	-    &$\>\>$	  K0V	 \\	
HD 136352 b &$\>\>$ 	5.2788 &$\>\>$ 		11.577 &$\>\>$	    0.0933 &	0.18 &$\>\>$	-    &$\>\>$	  G4V	 \\	
Gl 581 c &$\>\>$ 	5.406 &$\>\>$ 		12.9182 &$\>\>$	    0.073  &	0.07 &$\>\>$	0.31 &$\>\>$	  M2.5V	 \\	
\botrule
\end{tabular}}
\end{table}

\clearpage

\begin{table}
\caption{Continuing from Table 1. Currently known extrasolar planets with masses up to 10 Earth-masses. 
The quantities $M, P, a$ and $e$ represent the mass (in terms of Earth's mass $M_\oplus$), 
orbital period, semimajor axis, and orbital eccentricity of the planet. The mass 
of the central star is shown by $M_*$ and is given in the units of solar-masses $(M_\odot)$.}
{\begin{tabular}{@{}lllllll}
\toprule 
Planet & $M (M_\oplus$) & $P$ (day) & $a$ (AU)& 
$\>e$ & Stellar Type &  ${M_*} (M_\odot)$ \\
\colrule 
OGLE-2005-390L b &$\>\>$ 	5.406 &$\>\>$ 		3500  &$\>\>$	    2.1	   &	-    &$\>\>$	0.22 &$\>\>$	  M	 \\	
GJ 667C b      &$\>\>$ 	5.46324 &$\>\>$ 	7.199 &$\>\>$	    0.0504 &	0.09 &$\>\>$	0.33 &$\>\>$	  M1.5V	 \\	
GJ 433 b &$\>\>$ 	5.7876 &$\>\>$ 		7.3709 &$\>\>$	    0.058  &	0.08 &$\>\>$	0.48 &$\>\>$	  M1.5	 \\	
HD 1461 c &$\>\>$ 	5.9148 &$\>\>$ 		13.505 &$\>\>$	    0.1117 &	0    &$\>\>$	1.08 &$\>\>$	  G0V	 \\	
HD 39194 c &$\>\>$ 	5.9466 &$\>\>$ 		14.025 &$\>\>$	    0.0954 &	0.11 &$\>\>$	-    &$\>\>$	  K0V	 \\	
Gl 581 d &$\>\>$ 	6.042 &$\>\>$ 		66.64 &$\>\>$	    0.22   &	0.25 &$\>\>$	0.31 &$\>\>$	  M2.5V	 \\	
Kepler-11 d &$\>\>$ 	6.10242 &$\>\>$ 	22.68719&$\>\>$     0.159  &	0    &$\>\>$	0.95 &$\>\>$	  G	 \\	
HD 154088 b &$\>\>$ 	6.1374 &$\>\>$ 		18.596 &$\>\>$	    0.1316 &	0.38 &$\>\>$	-    &$\>\>$	  K0IV	 \\	
GJ 1214 b &$\>\>$ 	6.36 &$\>\>$ 		1.58040482 &$\>\>$  0.014  &	0.27 &$\>\>$	0.153 &$\>\>$	  M	 \\	
HD 215497 b &$\>\>$ 	6.36 &$\>\>$ 		3.93404 &$\>\>$	    0.047  &	0.16 &$\>\>$	0.87 &$\>\>$	  K3V	 \\	
HD 97658 b &$\>\>$ 	6.36 &$\>\>$ 		9.4957 &$\>\>$	    0.0797 &	0.13 &$\>\>$	0.85 &$\>\>$	  K1V	 \\	
Gl 876 d &$\>\>$ 	6.678 &$\>\>$ 		1.93778 &$\>\>$	    0.0208 &	0.21 &$\>\>$	0.334 &$\>\>$	  M4 V	 \\
HD 40307 c &$\>\>$ 	6.8688 &$\>\>$ 		9.62 &$\>\>$	    0.081  &	0    &$\>\>$	0.77 &$\>\>$	  K2.5V	 \\	
Kepler-18 b &$\>\>$ 	6.9006 &$\>\>$ 		3.504725 &$\>\>$    0.0447 &	-    &$\>\>$	0.972 &$\>\>$	  -	 \\	
GJ 3634 b &$\>\>$ 	6.996 &$\>\>$ 		2.64561 &$\>\>$	    0.0287 &	0.08 &$\>\>$	0.45 &$\>\>$	  M2.5	 \\	
Kepler-9 d &$\>\>$ 	6.996 &$\>\>$ 		1.592851 &$\>\>$    0.0273 &	-    &$\>\>$	1 &$\>\>$	  -	 \\	
HD 181433 b &$\>\>$ 	7.5684 &$\>\>$ 		9.3743 &$\>\>$	    0.08   &	0.39 &$\>\>$	0.78 &$\>\>$	  K3IV	 \\	
HD 1461 b &$\>\>$ 	7.6002 &$\>\>$ 		5.7727 &$\>\>$	    0.063  &	0.14 &$\>\>$	1.08 &$\>\>$	  G0V	 \\	
HD 93385 b &$\>\>$ 	8.3634 &$\>\>$ 		13.186 &$\>\>$	    0.1116 &	0.15 &$\>\>$	-    &$\>\>$	  G2V	 \\	
CoRoT-7 c &$\>\>$ 	8.3952 &$\>\>$ 		3.698 &$\>\>$	    0.046  &	0    &$\>\>$	0.93 &$\>\>$	  K0V	 \\	
Kepler-11 e &$\>\>$ 	8.40474 &$\>\>$ 	31.9959&$\>\>$	    0.194  &	0    &$\>\>$	0.95 &$\>\>$	  G	 \\	
GJ 176 b &$\>\>$ 	8.427 &$\>\>$ 		8.7836 &$\>\>$	    0.066  &	0    &$\>\>$	0.49 &$\>\>$	  M2.5V	 \\	
55 Cnc e &$\>\>$ 	8.586 &$\>\>$ 		0.7365449 &$\>\>$   0.0156 &	0.06 &$\>\>$	0.905 &$\>\>$	  K0IV-V \\	
Kepler-20 b &$\>\>$ 	8.586 &$\>\>$ 		3.6961219 &$\>\>$   0.0453 &	0.32 &$\>\>$	0.912 &$\>\>$	  G8	 \\	
HD 96700 b &$\>\>$ 	9.0312 &$\>\>$ 		8.1256 &$\>\>$	    0.0774 &	0.1  &$\>\>$	-     &$\>\>$     G0V	 \\	
HD 40307 d &$\>\>$ 	9.1584 &$\>\>$ 		20.46 &$\>\>$	    0.134  &	0    &$\>\>$	0.77 &$\>\>$	  K2.5V	 \\	
HD 7924 b &$\>\>$ 	9.222 &$\>\>$ 		5.3978 &$\>\>$	    0.057  &	0.17 &$\>\>$	0.832 &$\>\>$	  KOV	 \\	
HD 134606 b &$\>\>$ 	9.2856 &$\>\>$ 		12.083 &$\>\>$	    0.102  &	0.15 &$\>\>$	-     &$\>\>$	  G6IV	 \\	
HD 136352 d &$\>\>$ 	9.54 &$\>\>$ 		106.72 &$\>\>$	    0.411  &	0.43 &$\>\>$	-     &$\>\>$	  G4V	 \\	
HD 189567 b &$\>\>$ 	10.0488 &$\>\>$ 	-      &$\>\>$	    14.275 &    0.11 &$\>\>$	0.23 &$\>\>$	  G2V	 \\	
HD 93385 c &$\>\>$ 	10.1124 &$\>\>$ 	-      &$\>\>$	    46.025 &    0.21 &$\>\>$	0.24  &$\>\>$     G2V	 \\	
\botrule
\end{tabular}}
\end{table}

\clearpage
\begin{figure}
\centerline{\psfig{figure=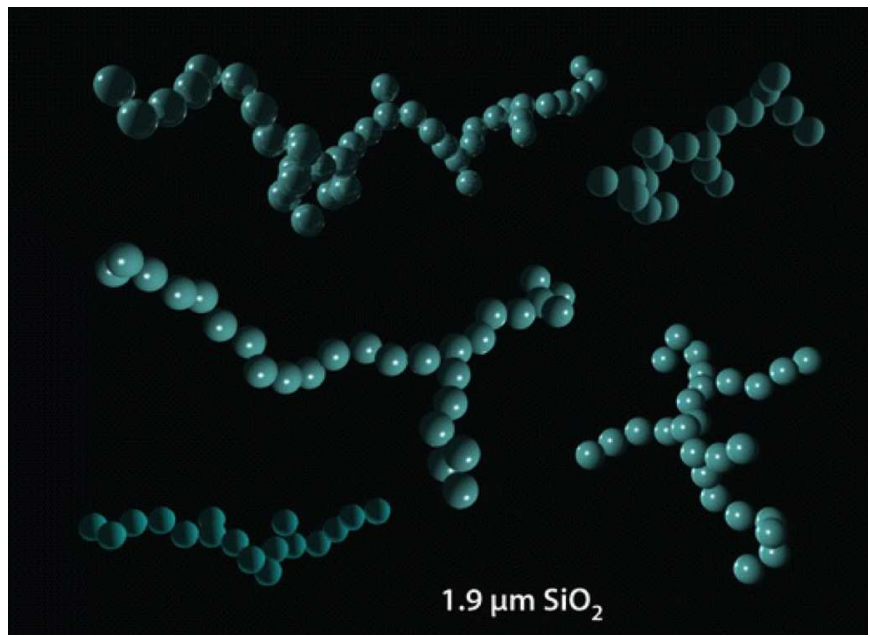,width=4.5in}}
\caption{
Coagulation of dust particles to fractal aggregates. Figure courtesy of J. Blum.}
\end{figure}

\clearpage
\begin{figure}
\centerline{\psfig{figure=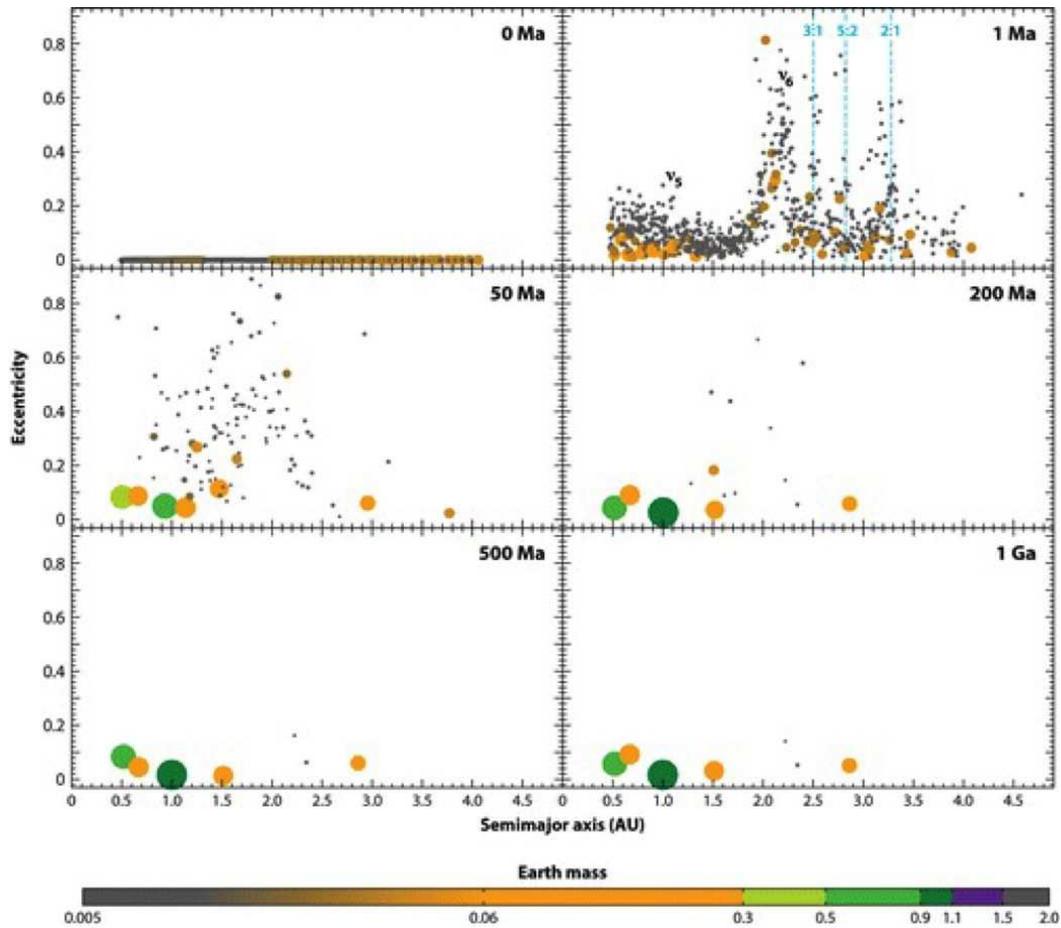,width=5.5in}}
\caption{Snapshots of the accretion of planetesimals and planetary embryos to terrestrial-size planets. 
The disk has a radial surface density profile of $-1.5$ with its value at 1 AU equal to 8 g cm$^{-3}$. 
Mean-motion and secular resonances with Jupiter and Saturn are also shown. Figure courtesy of A. Izidoro.}
\end{figure}

\clearpage
\begin{figure}
\centerline{\psfig{figure=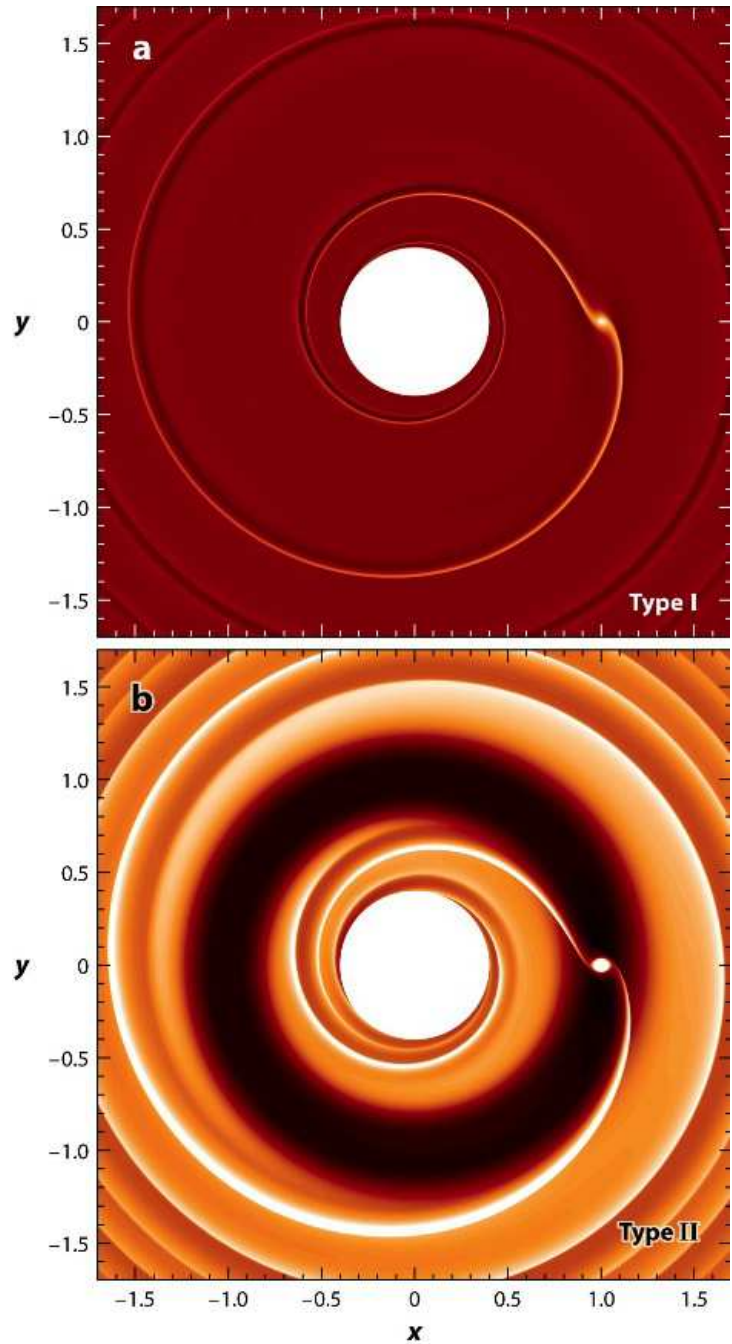,width=3.75in}}
\caption{Type I (top) and type II (bottom) planetary migration. Figures courtesy of F. Mass\'et.}
\end{figure}

\clearpage
\begin{figure}
\centerline{\psfig{figure=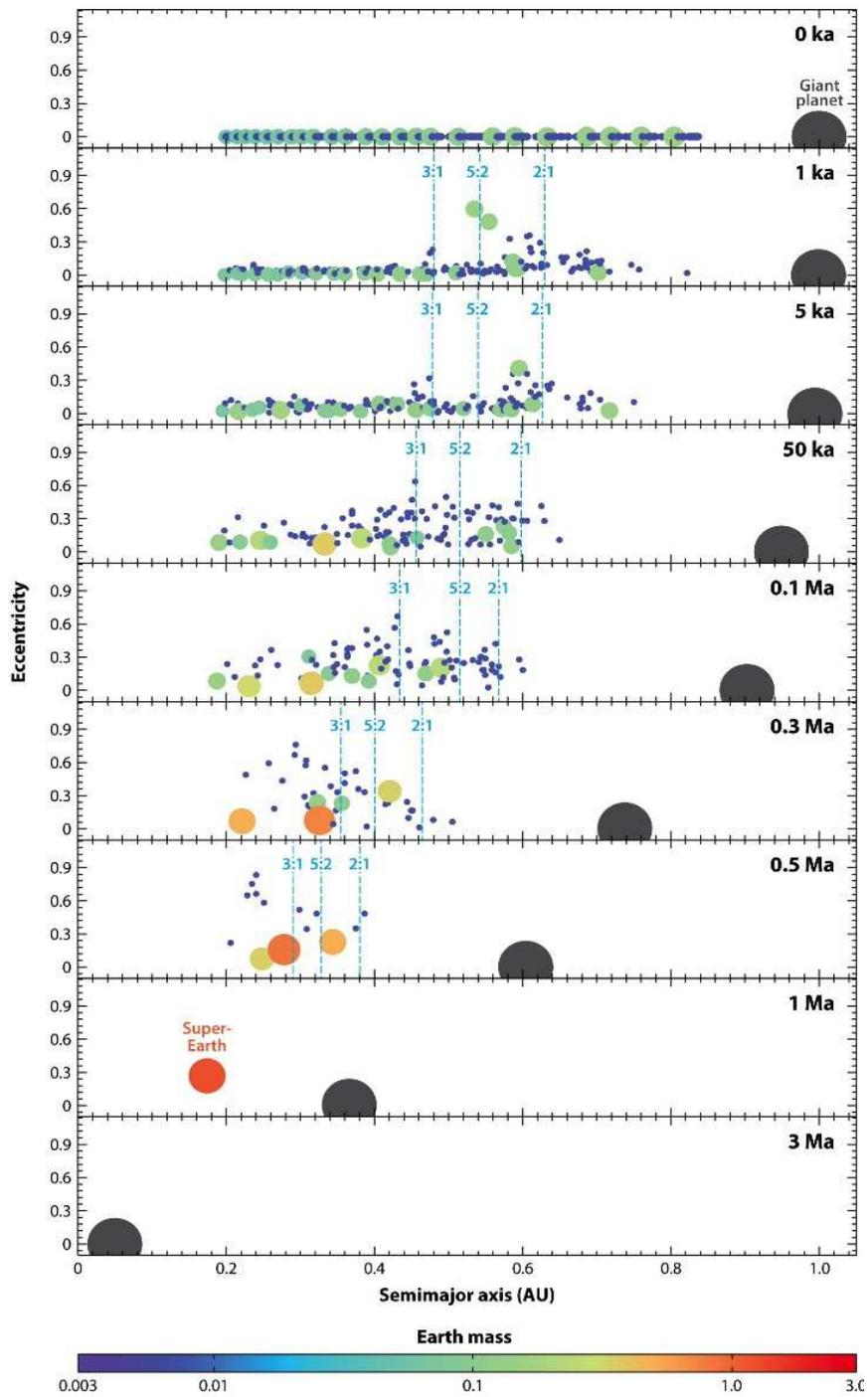,width=4.5in}}
\caption{Accretion of protoplanetary bodies during the migration of a giant planet
around a $0.3 {M_\odot}$ M star (Haghighipour \& Rastegar 2011).}
\end{figure}

\clearpage
\begin{figure}
\centerline{\psfig{figure=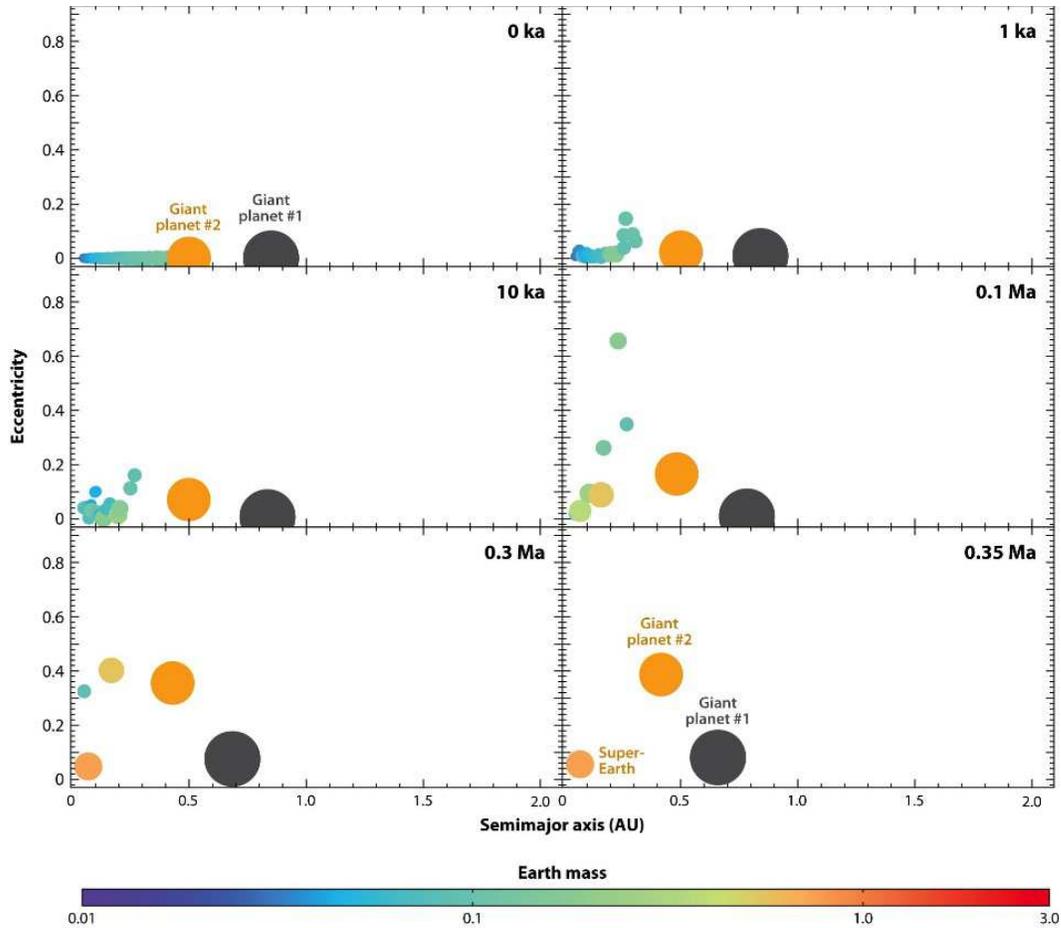,width=5.5in}}
\caption{Accretion of protoplanetary bodies during the migration of two giant planets around a 
$0.3 {M_\odot}$ M star. As shown here, the system becomes stable with two giant planets in a 1:2 MMR 
and a super-Earth in a short-period orbit (e.g., GJ 876).}
\end{figure}

\clearpage
\begin{figure}
\centerline{\psfig{figure=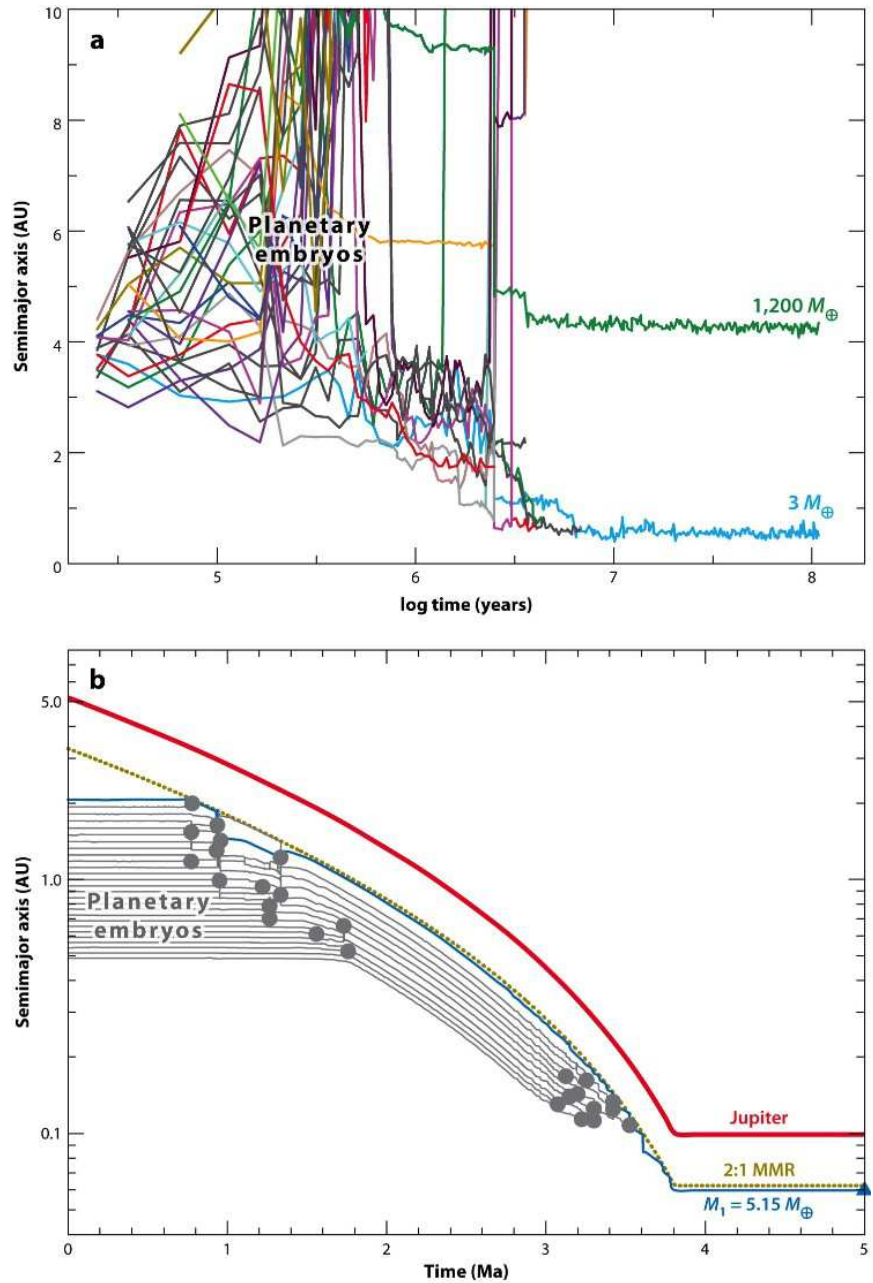,width=4.5in}}
\caption{Migration and accretion of planetary embryos and the formation of super-Earths. 
Top: The formation of an icy $3{M_\oplus}$ object at 0.5 AU. The super-Earth has two
giant companions, one at 10 AU (not shown here) and one at 4 AU with a mass of 1,200 $M_\oplus$. 
Figure courtesy of S. Kenyon. Bottom: A combination of the migration and accretion of embryos
to super-Earth bodies and their capture in MMR resonances. Figure courtesy of J.-L. Zhou.}

\end{figure}

\clearpage
\begin{figure}
\centerline{\psfig{figure=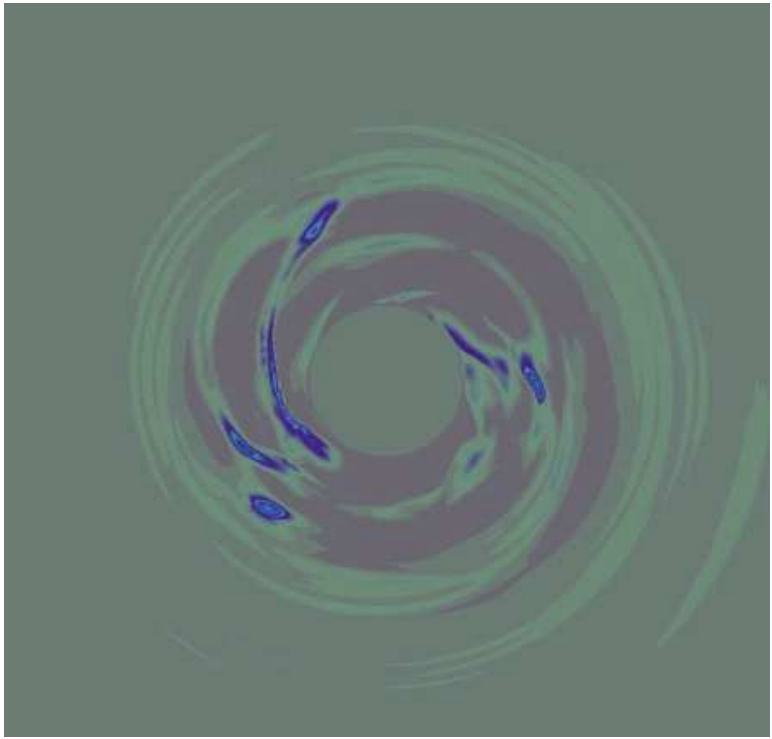,width=4in}}
\caption{A snapshot of a simulation of the formation of super-Earths around a 0.5 ${M_\odot}$
star in the disk-instability model. The four clumps shown in light blue are potential mini-Neptune 
and super-Earth objects. Figure courtesy of A. Boss.}
\end{figure}

\end{document}